\pdfoutput=1  

\documentclass[aps,prd,superscriptaddress,floatfix,nofootinbib,preprintnumbers,eqsecnum,twocolumn,dvipsnames]{revtex4-1}

\usepackage{amsmath,float,graphicx,placeins,amssymb,multirow,pgfplots,pgfplotstable}
\usepackage{empheq}
\usepackage{tikz}
\usepackage{comment}
\usepackage{enumerate,slashed,cancel,comment,color,bbm,graphicx,rotating,placeins,mathtools,multirow,hhline}
\usepackage[normalem]{ulem}
\usepackage{array}

\usepackage{hyperref}
\usepackage{color}
 \hypersetup
 {
   colorlinks,
   linkcolor={blue!80!black},
   citecolor={blue!70!black},
   urlcolor={blue!70!black}
 }

\usepackage{lipsum}
\usepackage{diagbox}
\usetikzlibrary{shapes.geometric,arrows}
\graphicspath{}

\newcommand{\expt}[1]{\left\langle #1 \right\rangle}
\newcommand{\abs}[1]{\left| #1 \right|}

\def\beq{\begin{equation}}
\def\eeq{\end{equation}}
\def\beqn{\begin{eqnarray}}
\def\eeqn{\end{eqnarray}}

\def\be{\begin{equation}}
\def\ee{\end{equation}}
\def\bea{\begin{eqnarray}}
\def\eea{\end{eqnarray}}
\def\half{\mbox{\small ${\frac{1}{2}}$}}

\def\quarter{\mbox{\small ${\frac{1}{4}}$}}

\newcommand{\newc}{\newcommand}

\def\calM{{\cal M}}
\def\calV{{\cal V}}
\def\calF{{\cal F}}

\def\bQ{{\bf Q}}
\def\bT{{\bf T}}

\def\Qs{{\bf q}}

\def\barOmega{{\overline{\Omega}}}

\def\barkappa{{\overline{\kappa}}}
\def\starbarkappa{{\overline{\kappa}^\ast}}
\def\barw{{\overline{w}}}
\def\BG{{\rm BG}}

\def\half{{\textstyle{1\over 2}}}
\def\quarter{{\textstyle{1\over 4}}}
\def\ie{{\it i.e.}\/}
\def\eg{{\it e.g.}\/}
\def\etc{{\it etc}.\/}
\def\inbar{\,\vrule height1.5ex width.4pt depth0pt}
\def\IR{\relax{\rm I\kern-.18em R}}
\def\SR{{\rm SR}}
\def\osc{{\rm osc}}

\font\cmss=cmss10 \font\cmsss=cmss10 at 7pt
\def\IQ{\relax{\rm I\kern-.18em Q}}
\def\IZ{\relax\ifmmode\mathchoice
 {\hbox{\cmss Z\kern-.4em Z}}{\hbox{\cmss Z\kern-.4em Z}}
 {\lower.9pt\hbox{\cmsss Z\kern-.4em Z}}
 {\lower1.2pt\hbox{\cmsss Z\kern-.4em Z}}\else{\cmss Z\kern-.4em Z}\fi}

\begin{document}

\title{
Cosmological Stasis from Dynamical Scalars:\\
Tracking Solutions and the Possibility of a Stasis-Induced Inflation\\
}

\def\andname{\hspace*{-0.5em}} 

\author{Keith R. Dienes}
 \email[Email address: ]{dienes@arizona.edu}
 \affiliation{Department of Physics, University of Arizona, Tucson, AZ 85721 USA}
 \affiliation{Department of Physics, University of Maryland, College Park, MD 20742 USA}
\author{Lucien Heurtier}
\email[Email address: ]{lucien.heurtier@kcl.ac.uk}
\affiliation{Theoretical Particle Physics and Cosmology, King’s College London,
Strand, London WC2R 2LS, United Kingdom}
\author{Fei Huang}
\email[Email address: ]{fei.huang@weizmann.ac.il}
 \affiliation{Department of Particle Physics and Astrophysics, Weizmann Institute of Science, Rehovot 7610001, Israel}
  \author{Tim M.P. Tait}
 \email[Email address: ]{ttait@uci.edu}
 \affiliation{Department of Physics and Astronomy, University of California, Irvine, CA  92697  USA}
 \author{Brooks Thomas}
 \email[Email address: ]{thomasbd@lafayette.edu}
 \affiliation{Department of Physics, Lafayette College, Easton, PA  18042 USA}

%

\begin{abstract}
It has recently been realized that many theories of physics beyond the 
Standard Model give rise to cosmological histories exhibiting extended epochs 
of {\it cosmological stasis}\/.  During such epochs, the abundances of different 
energy components such as matter, radiation, and vacuum energy each remain fixed
despite cosmological expansion.  In previous analyses of the stasis phenomenon, 
these different energy components were modeled as fluids with fixed, unchanging 
equations of state.   In this paper, by contrast, we consider more realistic systems 
involving dynamical scalars which pass through underdamping transitions as the universe 
expands.   Indeed, such systems might be highly relevant for BSM scenarios involving 
higher-dimensional bulk moduli and inflatons.  Remarkably, we find that stasis emerges 
even in such situations, despite the appearance of time-varying equations of state.   
Moreover, this stasis includes several new features which might have important 
phenomenological implications and applications.  For example, in the presence of an 
additional ``background'' energy component, we find that the scalars evolve into a 
``tracking'' stasis in which the stasis equation of state automatically tracks that of 
the background.  This phenomenon exists even if the background has only a small initial 
abundance.  We also discuss the intriguing possibility that our results
might form the basis of a new ``Stasis Inflation'' scenario in which no {\it ad-hoc}\/ 
inflaton potential is needed and in which there is no graceful-exit problem.  Within 
such a scenario, the number of $e$-folds of cosmological expansion produced is directly related to 
the hierarchies between physical BSM mass scales.  Moreover, non-zero matter and 
radiation abundances  can be sustained throughout the inflationary epoch.
\end{abstract}
\maketitle

\tableofcontents

\def\ie{{\it i.e.}\/}
\def\eg{{\it e.g.}\/}
\def\etc{{\it etc}.\/}
\def\taubar{{\overline{\tau}}}
\def\qbar{{\overline{q}}}
\def\kbar{{\overline{k}}}

\def\beq{\begin{equation}}
\def\eeq{\end{equation}}
\def\beqn{\begin{eqnarray}}
\def\eeqn{\end{eqnarray}}
\def\apo{\mbox{\small ${\frac{\alpha'}{2}}$}}
\def\half{\mbox{\small ${\frac{1}{2}}$}}
\def\sqapo{\mbox{\tiny $\sqrt{\frac{\alpha'}{2}}$}}
\def\sqap{\mbox{\tiny $\sqrt{{\alpha'}}$}}
\def\sqapxtwo{\mbox{\tiny $\sqrt{2{\alpha'}}$}}
\def\aptwo{\mbox{\tiny ${\frac{\alpha'}{2}}$}}
\def\apofour{\mbox{\tiny ${\frac{\alpha'}{4}}$}}
\def\bosqtwo{\mbox{\tiny ${\frac{\beta}{\sqrt{2}}}$}}
\def\btosqtwo{\mbox{\tiny ${\frac{\tilde{\beta}}{\sqrt{2}}}$}}
\def\apofour{\mbox{\tiny ${\frac{\alpha'}{4}}$}}
\def\sqaptwo{\mbox{\tiny $\sqrt{\frac{\alpha'}{2}}$}  }
\def\apoeight{\mbox{\tiny ${\frac{\alpha'}{8}}$}}
\def\sapoeight{\mbox{\tiny ${\frac{\sqrt{\alpha'}}{8}}$}}

\newc{\gsim}{\lower.7ex\hbox{{\mbox{$\;\stackrel{\textstyle>}{\sim}\;$}}}}
\newc{\lsim}{\lower.7ex\hbox{{\mbox{$\;\stackrel{\textstyle<}{\sim}\;$}}}}
\def\calM{{\cal M}}
\def\calV{{\cal V}}
\def\calF{{\cal F}}
\def\bQ{{\bf Q}}
\def\bT{{\bf T}}
\def\Qs{{\bf q}}

\def\half{{\textstyle{1\over 2}}}
\def\quarter{{\textstyle{1\over 4}}}
\def\ie{{\it i.e.}\/}
\def\eg{{\it e.g.}\/}
\def\etc{{\it etc}.\/}
\def\inbar{\,\vrule height1.5ex width.4pt depth0pt}
\def\IR{\relax{\rm I\kern-.18em R}}
 \font\cmss=cmss10 \font\cmsss=cmss10 at 7pt
\def\IQ{\relax{\rm I\kern-.18em Q}}
\def\IZ{\relax\ifmmode\mathchoice
 {\hbox{\cmss Z\kern-.4em Z}}{\hbox{\cmss Z\kern-.4em Z}}
 {\lower.9pt\hbox{\cmsss Z\kern-.4em Z}}
 {\lower1.2pt\hbox{\cmsss Z\kern-.4em Z}}\else{\cmss Z\kern-.4em Z}\fi}
\def\trho{\tilde{\rho}}
\def\hatt{\hat{t}}
\def\tt{\tilde{t}}
\def\nn{\nonumber}


\section{Introduction, motivation, and overview of results}


Cosmological stasis~\cite{Dienes:2021woi,Dienes:2023ziv} is a surprising phenomenon wherein 
the abundances of multiple cosmological energy components (\eg, matter, radiation, 
vacuum energy, \etc)\  with different equations of state each remain constant over an 
extended period, despite the effects of Hubble expansion.  This phenomenon has been shown 
to arise in new-physics scenarios involving towers of unstable particles~\cite{Dienes:2021woi}, 
theories involving populations of scalars undergoing underdamping transitions~\cite{Dienes:2023ziv}, 
and even theories with populations of primordial black holes with extended mass
spectra~\cite{Barrow:1991dn,Dienes:2022zgd}. 

In all of these realizations of stasis, the energy densities of the different energy 
components involved scale differently under cosmological expansion because they have 
different equations of state.  Thus, {\it a priori}\/, one might expect their 
respective abundances to change rapidly as the universe expands.  However, these 
changes in the abundances of the different energy components can be compensated by 
processes that actually transfer energy between these different components.  
In this way, each of the different abundances can potentially remain constant.  

At first glance, it might seem that one must carefully balance the effects of these 
energy-transferring processes against the effects of cosmological expansion in order to 
achieve stasis.  If true, this would render  stasis the result of a severe fine-tuning. 
However, as shown in Refs.~\cite{Dienes:2021woi,Dienes:2022zgd,Dienes:2023ziv,Barrow:1991dn}, 
the required balancing is actually a global attractor within the coupled system of Boltzmann 
and Einstein equations that govern the cosmological evolution of the abundances.
The universe will thus necessarily evolve towards (and eventually enter) stasis 
irrespective of initial conditions.

There exist many different examples of such energy-transferring processes.  These in turn 
depend on the particular model of stasis under study.   For example, in models exhibiting a 
stasis between particulate matter and radiation, as discussed in 
Refs.~\cite{Dienes:2021woi,Dienes:2023ziv}, the relevant energy-transferring process was 
particle decay.   Likewise, in models of matter/radiation stasis in which the matter takes 
the form of primordial black holes~\cite{Barrow:1991dn,Dienes:2022zgd}, the relevant 
energy-transferring process was Hawking evaporation.  Indeed, both particle decay and 
Hawking radiation convert matter energy to radiation energy and therefore play an integral 
role in keeping the abundances of matter and radiation constant despite cosmological expansion.

In Ref.~\cite{Dienes:2023ziv}, by contrast, it was shown that stasis can also arise between 
vacuum energy and matter.   In fact, it was even shown that one can have 
a {\it triple}\/ stasis between vacuum energy, matter, and radiation 
simultaneously~\cite{Dienes:2023ziv}.  The underlying model that was analyzed for these 
purposes was built upon the dynamical evolutions of the homogeneous zero-mode field values 
associated with a tower of scalar fields.  As is well known, each such field value evolves 
according to an equation of motion which resembles that of a massive harmonic oscillator 
with a Hubble-induced ``friction'' term.   Within an expanding universe, the Hubble-friction 
term is large at early times, and thus our field is overdamped.  In this case, the potential 
energy of the field vastly exceeds its kinetic energy, whereupon the energy density of this 
field may be viewed as pure potential energy (\ie, vacuum energy), with an equation-of-state 
parameter $w\approx -1$.  However, as the universe continues to expand, the Hubble 
parameter drops, eventually reaching (and passing through) a critical point at which our
system becomes underdamped and our field begins to oscillate and eventually virialize.  
During such an oscillatory phase, the kinetic and potential energies associated with our 
field are then approximately equal, whereupon we find that $w\approx 0$.  This transition 
from an overdamped phase to an underdamped phase may thus be regarded as an energy-transferring 
process in which the corresponding energy density transitions from vacuum energy to matter.

In each of these previous realizations of the stasis phenomenon, the corresponding energy 
densities were modeled as fluids with {\it time-independent}\/ equations of state.  Indeed, 
in cases involving stases between matter and radiation --- such as were discussed in
Refs.~\cite{Dienes:2021woi,Dienes:2023ziv} --- the matter and radiation were modeled as 
fluids with constant equation-of-state parameters $w=0$ and $w=1/3$, respectively.  Given 
that the physics in these cases rested on either particle decay or Hawking radiation, 
this can be viewed as a natural and reasonable assumption.  

For calculational simplicity, the same assumption was also made in Ref.~\cite{Dienes:2023ziv} 
when considering the dynamical evolution of the homogeneous zero-mode field value associated 
with a scalar field.  In particular, the energy density associated with such a field was treated 
in Ref.~\cite{Dienes:2023ziv} as that of a fluid with a constant equation-of-state parameter 
$w=0$ throughout the later, underdamped phase, and treated as that of a fluid with a constant 
equation-of-state parameter $w$ near $-1$ throughout the earlier, overdamped phase. 
Moreover, the transition between these two phases of the theory was treated as instantaneous, 
occurring at the critical time at which the underdamping transition normally takes place 
in the fully dynamical theory.  Given these assumptions, it was then found that a stasis also 
emerged between these two fluids --- a stasis which was interpreted as existing between vacuum 
energy and matter.  Moreover, as noted above, allowing these fields to decay after transitioning 
from vacuum energy to matter was then shown to result in a triple stasis between vacuum 
energy, matter, and radiation.

While such results are exciting and may have many phenomenological implications, 
the true dynamical evolution of a scalar field in a cosmological setting is more complicated 
than this.  As noted above, the true dynamics of such a field is governed by an equation of 
motion which is that of a damped harmonic oscillator.   Within such a system, there continues 
to exist a critical boundary between an overdamped and underdamped phase as the Hubble-friction 
term decreases over time.  However, the equation-of-state parameter prior to this transition 
is not fixed at a small value near $w\approx -1$ within the overdamped phase, nor is it 
(or its virial time-average) fixed at $w=0$ within the subsequent underdamped phase.   
Instead, the true behavior of our dynamical scalar field is an entirely smooth one.  
The corresponding equation-of-state parameter will indeed {\it asymptote}\/ to a fixed value 
near $w= -1$  at increasingly early times --- an epoch during which the corresponding 
field $\phi_\ell$ remains fixed or at most slowly rolls --- 
and likewise it will  {\it asymptote}\/ to the fixed time-averaged value $w=0$  at increasingly 
late times, an epoch during which the field experiences rapid virialized oscillations.   However, 
between these asymptotic limits, our scalar field and its equation of state are both evolving 
dynamically in a smooth, non-trivial, time-dependent manner.  This evolution does not even 
exhibit a sharp change of behavior of any sort as our system passes through the critical 
underdamping transition.

In this paper, we seek to determine what happens to our stasis phenomenon when we take 
this full time-dependence into account.  At first glance, it might seem that including 
this time dependence for the equation-of-state parameters for the individual scalar fields
would completely destabilize the stasis that emerges when these equation-of-state parameters 
are instead taken to be constant both before and after the underdamping transition.   
Indeed, such time-dependent equation-of-state parameters could in principle 
complicate the manner in which the Hubble parameter evolves with time, and thus 
lead to a more complicated dynamical evolution for our scalar fields and their 
corresponding energy densities.  However, since stasis is built on the idea that the 
abundances of our different fluids remain constant despite cosmological expansion, it 
is a natural expectation that allowing for time-varying equation-of-state parameters would 
destroy the stasis that is observed when these equation-of-state parameters are constant.

Remarkably, in this paper we find that stasis can emerge even in such situations.  
In particular, we find that there exists a large class of scenarios in which stasis 
emerges as an attractor --- the time-variation of the equation-of-state parameters
for the individual fields notwithstanding ---
and persists across many $e$-folds of cosmological expansion.  In this regard, then, 
our results are similar to those of Refs.~\cite{Dienes:2021woi, Dienes:2023ziv} and 
demonstrate that the stasis phenomenon exists even for dynamical scalars when their 
full time dependence is taken into account.

Despite this similarity, we shall find that the stasis that is realized through 
fully dynamical scalars has a number of additional properties that transcend those 
arising within the previous realizations of stasis which have been identified.
In particular, we shall find that the fundamental constraint equation that underlies 
this stasis does not uniquely predict the equation of state of our system once it has 
entered stasis.  This gives our dynamical-scalar stasis a certain intrinsic mathematical 
flexibility that was not previously available.

As we shall find, the full implications of this additional flexibility are particularly 
significant when this stasis is realized in the presence of an additional  
``background spectator'' fluid --- \ie, a fluid which is completely inert, neither 
receiving energy from our stasis system nor donating energy to it.  Indeed, regardless 
of the initial abundance and equation-of-state parameter which are assumed for this 
background fluid, we find not only that our stasis solution continues to exist, but 
that it actually has the flexibility needed in order to  {\it track}\/ this fluid, 
automatically adjusting its properties such that the resulting equation-of-state 
parameter $w_{\rm univ}$ for the universe as a whole during stasis matches that of 
the background.  This is thus our first example of a ``tracking'' stasis.   
Indeed, we find that this tracking property persists even if the equation of state 
for the background fluid changes with time. 

This realization of stasis involving dynamical scalars also has another important 
property: as we shall see, it can easily accommodate an equation-of-state parameter 
for the universe within the range $-1 < w_{\rm univ} < -1/3$.  A stasis epoch in
which $w_{\rm univ}$ falls within this range constitutes a period of {\it accelerated}\/ 
cosmological expansion in which $\ddot a>0$, where $a$ is the scale factor.  
Since a stasis epoch of this sort can span many $e$-folds of expansion, such a
epoch can serve as a means of addressing the horizon and flatness problems.  This 
observation suggests that stasis could potentially serve as a novel mechanism for 
achieving cosmic inflation.  No non-trivial, {\it ad-hoc}\/ inflaton potential would be 
required within such a ``Stasis Inflation'' scenario; likewise, this scenario has no 
graceful-exit problem.  Moreover, non-zero matter and radiation abundances can be 
sustained throughout an inflationary epoch of this sort.  In this paper, we shall 
discuss this new ``Stasis Inflation'' possibility and outline some of its key 
qualitative features.  Of course, further analysis will be required in order to 
determine whether such an inflation scenario is truly viable.

This paper is organized as follows.  In Sect.~\ref{sec:single_scalar}, we review the 
dynamical evolution of a single scalar field which undergoes a transition from 
overdamped rolling to underdamped oscillation.  In Sect.~\ref{sec:ScalarTower}, 
we then extend this single-field analysis to the more general case in which the 
particle content of the theory includes a tower of scalar fields $\phi_\ell$ with a 
non-trivial spectrum of masses and initial abundances.  Despite the non-trivial 
manner in which the individual equation-of-state parameters $w_\ell(t)$ for these 
scalars each evolve in time, we nevertheless find that the tower as a whole can 
give rise to a stasis epoch in which the effective equation-of-state parameter 
for the universe as a whole is essentially constant.  In Sect.~\ref{sec:tracker}, 
we then consider how the resulting cosmological dynamics is modified in the presence 
of an additional background energy component with an equation-of-state parameter 
$w_{\rm BG}$.  We find that the tower of scalars can still reach a stasis.  
In fact, for certain values of $w_{\rm BG}$, we find that the equation-of-state 
parameter for the tower evolves toward $w_{\rm BG}$ and tracks it, even in situations 
in which $w_{\rm BG}$ exhibits a non-trivial time-dependence.  
In Sect.~\ref{sec:inflation}, we then discuss the possibility of a stasis-induced 
inflationary epoch during which our stasis itself drives an accelerated expansion 
of the universe.  Finally, in Sect.~\ref{sec:Conclusions}, we conclude with a 
summary of our main results and a discussion of possible avenues for future work.


\FloatBarrier
\section{A single scalar in an expanding universe\label{sec:single_scalar}}


Let us start by reviewing the dynamical evolution of a single real scalar field 
$\phi$ in an expanding universe.  In general, the energy density and pressure 
of such a scalar field are given by
\beq
  \rho_\phi~=~\frac{1}{2}\dot\phi^2+V\,,~~~ 
    P_\phi~=~\frac{1}{2}\dot\phi^2-V\,,
    \label{eq:rho_pressure}
\eeq
where the ``dot'' denotes the derivative with respect to the time $t$ in the
cosmological background frame and where $V(\phi)$ is the scalar potential.  
The equation-of-state parameter for this field is therefore thus given by
\beq
  w_\phi~\equiv~\frac{P_\phi}{\rho_\phi}~=~
    \frac{\frac{1}{2}\dot\phi^2-V}{\frac{1}{2}\dot\phi^2+V}~.
\label{eq:eos_phi}
\eeq
In general, $w_\phi$ is time-dependent and can vary continuously within the 
range $-1 \leq w_\phi \leq 1$.

The dynamics of this scalar field is governed by its equation of motion
\beq
  \ddot\phi+3H\dot\phi + \frac{dV}{d\phi}~=~0\,,
\label{eq:EOM_general}
\eeq
where the effects of the Friedmann-Robertson-Walker cosmology (\ie, the effects coming from the 
expansion of the universe) are encoded within the time-dependence of the Hubble 
parameter $H\equiv \dot a/a$.  As evident from 
Eq.~(\ref{eq:EOM_general}), the Hubble parameter affects the evolution of the 
scalar by providing a source of ``friction'' which damps the motion of the scalar.
The value of $H$ is related to the total energy density $\rho_{\rm tot}$  of the 
universe through the Friedmann equation
\beq
  H^2~=~\frac{8\pi G}{3}\rho_{\rm tot}~=~\frac{\rho_{\rm tot}}{3M_P^2}\,,
\label{eq:Friedmann_eqn}
\eeq
where $M_P=1/\sqrt{8\pi G}$ is the reduced Planck mass.  A larger value of
$\rho_{\rm tot}$ therefore corresponds to a larger damping term for $\phi$ and 
vice versa.  Moreover, in this paper we shall also make the ``minimal'' assumption 
that the potential is quadratic --- \ie, that 
\beq
  V(\phi)~=~\frac{1}{2}m^2\phi^2~,
\eeq
where $m$ is the mass of $\phi$.

In general, the solutions to Eq.~(\ref{eq:EOM_general}) will depend critically on the 
size of the Hubble-friction term.  When this term is sufficiently small, the system is 
underdamped and the value of $\phi$ oscillates with a decreasing amplitude.  By contrast, 
if the Hubble-friction term is sufficiently large, the system is overdamped and $\phi$ 
either remains effectively constant or decreases slowly without oscillating.
However, within a given cosmology, $H(t)$ generally decreases as a function of $t$.  
Thus, even if $\phi$ is initially in the overdamped phase, it will eventually transition 
to the underdamped phase when $H(t)$ drops below the critical value $H(t) = 2m/3$.

As a result of these features, it is of great interest to understand how $\rho_{\rm tot}$ 
(and therefore $H$) varies with time.  In particular, we shall focus on two cases of 
interest which represent different possible relationships between $\rho_\phi$ and 
$\rho_{\rm tot}$:
\begin{itemize}
  \item \uline{Case I}: In addition to $\phi$, the universe contains another cosmological
    energy component with a constant equation-of-state parameter $w$.  This additional 
    energy component is assumed to dominate the energy density of the universe during the 
    time period of interest.  Since $\rho_\phi \ll \rho_{\rm tot}$, the evolution of $H$ is 
    essentially independent of $\rho_\phi$ throughout this time period.
  \item \uline{Case II}: The field $\phi$ is the only cosmological energy component with 
    non-negligible energy density.  Thus, to a good approximation, we may take 
    $\rho_{\rm tot} = \rho_\phi$.
\end{itemize}
Both of these cases have been studied extensively, and we shall review the cosmological
dynamics which emerges in each case in turn.

\subsection{Case I: ~Fixed external cosmology\label{sec:fixed_background}}

\begin{figure}
\includegraphics[width=0.48\textwidth]{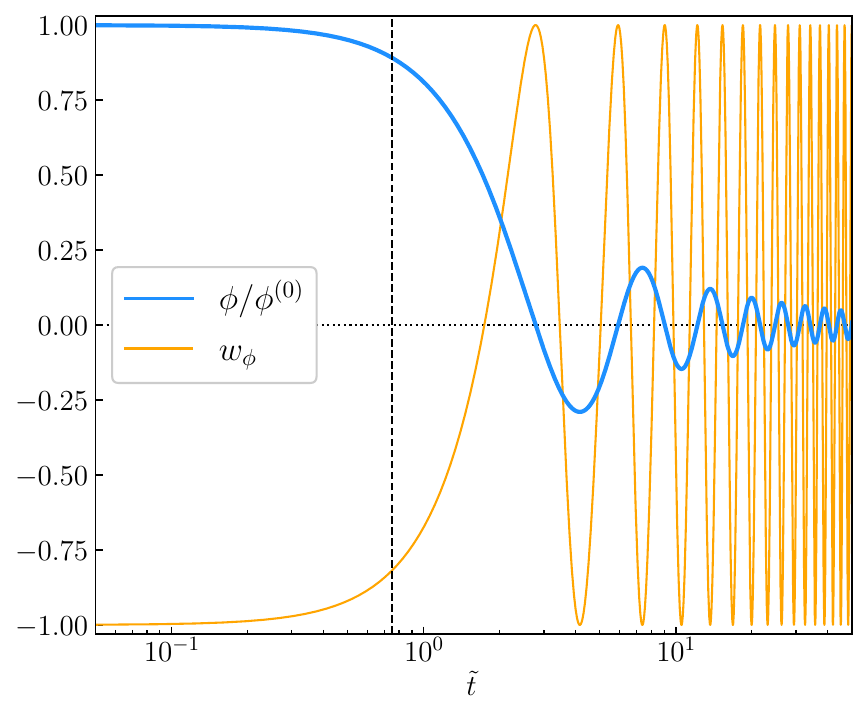}
\caption{The value of the scalar field $\phi(\tt)$, normalized to its asymptotic 
  early-time value $\phi^{(0)}$ and plotted as a function of the dimensionless 
  time variable $\tt$ during an epoch in which the energy density of the universe 
  is dominated by a radiation component ($w=1/3$).  Also shown is the corresponding 
  equation-of-state parameter $w_\phi(\tt)$.  The curves shown here correspond 
  to a choice of initial conditions in which $\phi^{(0)} \neq 0$ and $\phi'(\tt^{(0)})=0$.  
  The vertical dashed line at $\tt=\kappa/2$ indicates the critical value 
  $\tt_c$ of $\tt$ associated with the transition from overdamped to underdamped evolution.
    \label{fig:phi_t}}
\end{figure}

During any epoch wherein the universe is dominated by a cosmological energy 
component with a fixed equation-of-state parameter $w$, the Hubble parameter is  
given by $H = \kappa/(3t)$ with $\kappa=2/(1+w)$.  It therefore follows that
in Case~I, Eq.~\eqref{eq:EOM_general} reduces to
\beq
  \phi''+\frac{\kappa}{\tt}\phi' + \phi~=~0\,,
  \label{eq:EOM_reduced}
\eeq
where $\tt\equiv mt$ is a dimensionless time variable and where a prime denotes 
a derivative with respect to $\tt$.  The general solution to this differential 
equation takes the form
\beq
  \phi(\tt) ~=~ \tt^{(1-\kappa)/2}
  \left[ c_J J_{(\kappa-1)/2}(\tt)+c_Y Y_{(\kappa-1)/2}(\tt)\right]\,,
  \label{eq:phi_bessel}
\eeq
where $J_\nu(z)$ and $Y_\nu(z)$ are Bessel functions of the first and second kind, 
respectively, and where $c_J$ and $c_Y$ are coefficients with dimensions of mass.
This solution for $\phi(\tt)$ is plotted as a function of $\tt$ in Fig.~\ref{fig:phi_t}.

It is also possible to obtain an approximate solution for $\tilde t\ll 1$.
For $z \ll 1$, the Bessel functions $J_\nu(z)$ and $Y_\nu(z)$ are well 
approximated by $J_\nu(z)\sim z^\nu$ and $Y_\nu(z)\sim -z^{-\nu}$.
Thus, if the initial conditions for $\phi$ at some early dimensionless 
time $\tt^{(0)} \ll 1$ are such that $\phi^{(0)} \equiv \phi(\tt^{(0)})\not=0$ 
and $\phi'(\tt^{(0)}) \approx 0$, one may take $c_Y\approx 0$
and thereby obtain the approximate solution
\beq
  \phi(\tt) ~\approx~ c_J \,\tt^{(1-\kappa)/2}J_{(\kappa-1)/2}(\tt)\,. 
  \label{eq:phi_approx}
\eeq
This expression provides an excellent approximation to the full numerical solution 
for $\phi(\tt)$ shown in Fig.~\ref{fig:phi_t}.  Indeed, a plot of this approximate 
expression would be indistinguishable to the naked eye from the full solution 
over the entire range of $\tt$ shown.
 
As can be seen from Fig.~\ref{fig:phi_t}, the expression for $\phi(\tt)$ in 
Eq.~\eqref{eq:phi_bessel} behaves like a damped oscillator.  At early times, when 
$\tt \ll 1$, the field is overdamped due to the sizable Hubble-friction term.  
As a result,  we find that $w_\phi \approx -1$ within this regime, and $\phi$ 
behaves like a vacuum-energy component.  By contrast, at late times, when $\tt \gg 1$, 
both $\phi$ itself and $w_\phi$ oscillate rapidly.  The amplitude of $\phi$ decreases
with $\tt$ within this regime, and as a result we find  $\rho_\phi \sim a^{-3}$, 
just as we would expect for the energy density of massive matter.   Accordingly, while 
the {\it amplitude}\/ of $w_\phi$ is effectively unity within this regime, 
the {\it time-averaged}\/ value $\langle w_\phi\rangle_t$ of $w_\phi$ over a sufficiently 
long interval $\Delta \tt \equiv \tt - \tt^{(0)}$ approaches 
$\langle w_\phi\rangle_t \approx 0$.  

The transition region between these two limiting 
regimes physically corresponds to the time window wherein $\phi$ is rolling down its 
potential $V(\phi)$ with non-negligible field velocity $\phi'$, but has not yet reached 
the potential minimum at $\phi = 0$.  During this window, both $\phi$ and $w_\phi$  
evolve non-trivially with $\tt$.  Since this transition from overdamped and underdamped 
evolution occurs when $3H(t) \sim 2m$, as discussed above, it is conventional to define 
the critical time $t_c$ associated with this phase transition such that $\tt_c=\kappa/2$
in this case.  However, this phase transition clearly is not instantaneous, and as 
we shall see, the manner in which scalar fields evolve during such transition windows 
has important implications for the cosmological dynamics which emerges when a 
tower of such scalars is present.

\FloatBarrier
\subsection{Case II: ~Scalar domination\label{sec:scalar_domination}}

The cosmological dynamics which governs the evolution of $\phi$ and $H$ is 
significantly more complicated in Case~II than in Case~I due to
the fact that $H$ now depends on $\phi$ itself.  Indeed, in this case we find that 
Eq.~\eqref{eq:EOM_general} takes the form
\beq
  \phi''+\frac{3H}{m}\phi' + \phi ~=~ 0\,,
  \label{eq:EOM_reduced_2}
\eeq
where from Eq.~(\ref{eq:Friedmann_eqn}) we now have
\beq
  H~=~\frac{m}{\sqrt{6}}
    \sqrt{\left(\frac{\phi}{M_P}\right)^2+\left(\frac{\phi'}{M_P}\right)^2}\,.
  \label{eq:Hubble_CaseII}
\eeq
This dependence of the Hubble parameter on $\phi$ and $\phi'$ not only changes the 
time-evolution of $\phi$, but also introduces an added sensitivity of our system 
to its initial conditions.  For example, changing the initial value of $\phi$ has 
the effect of changing the initial value of the Hubble parameter $H$, and as we 
shall see, this can in turn affect the length of time that must elapse before 
our system can reach critical milestones such as the transition to an underdamped 
phase.

\begin{figure*}
\includegraphics[width=0.48\textwidth]{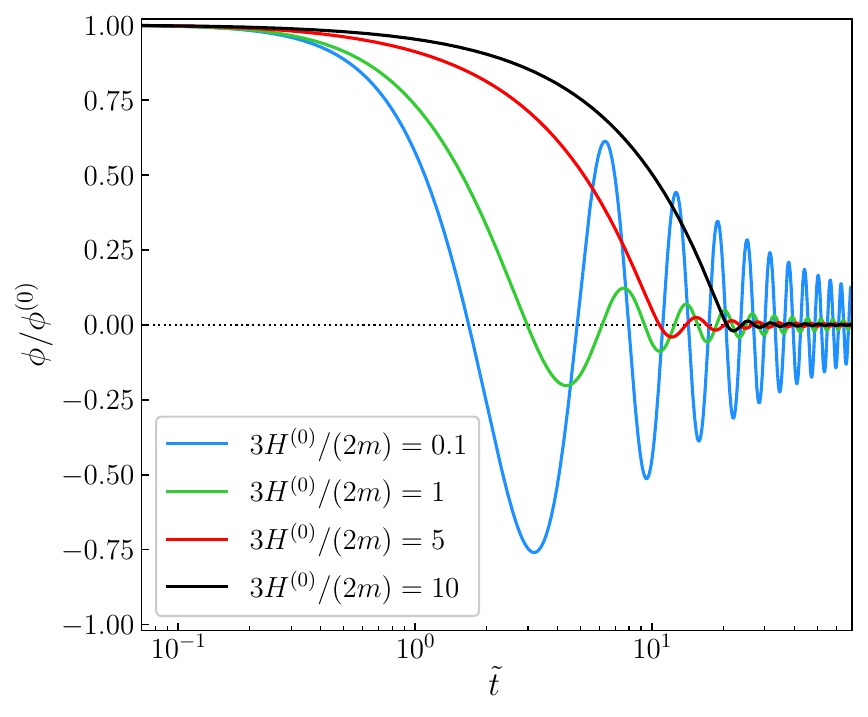}
\includegraphics[width=0.48\textwidth]{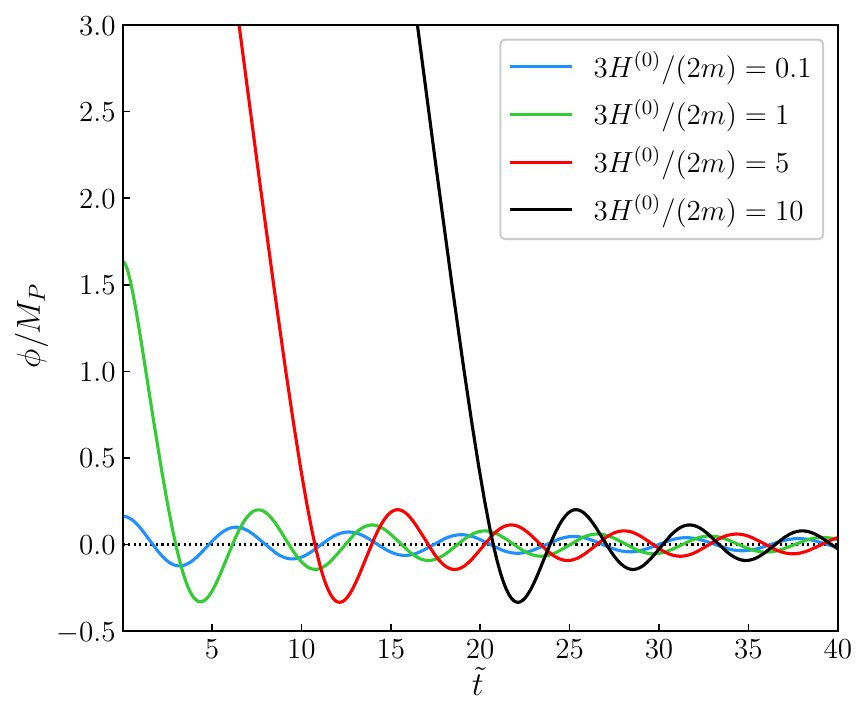}
\caption{{\it Left panel}\/: The values of the scalar field $\phi(\tt)$, normalized 
  to their asymptotic early-time values $\phi^{(0)}$ and  plotted as functions of the 
  dimensionless time variable $\tt$ for a variety of different choices of $\phi^{(0)}$ 
  during an epoch in which the energy density of the universe is dominated by $\phi$ itself.  
  Note that varying $\phi^{(0)}$ for the same value of $m$ is tantamount to choosing different 
  values of $3H^{(0)}/(2m)$.  In all cases, we have taken $\phi'(t^{(0)})=0$.  
  {\it Right panel}\/: Same as in the left panel, but with $\phi$ normalized to the 
  value of $M_P$ and with a linear rather than a logarithmic scale for the horizontal 
  axis.  Note that unlike the other curves, the blue curve always already begins in 
  the underdamped regime.  
\label{fig:phi_t_dom}}
\end{figure*}

Solutions to the non-linear differential equation in Eq.~(\ref{eq:EOM_reduced_2}) 
may be obtained numerically.  In examining the behavior of these solutions,
we once again focus for simplicity on the case in which the initial conditions 
for $\phi$ at $\tt^{(0)} \ll 1$ are such that $\phi^{(0)}$ is non-vanishing, 
while $\phi'(\tt^{(0)}) \approx 0$.  In Fig.~\ref{fig:phi_t_dom}, we show how 
$\phi(\tt)$ evolves as a function of $\tt$ for several different values of 
$3H^{(0)}/2m$ --- or, equivalently, since $3H^{(0)}/2m = \sqrt{3/8}\,|\phi^{(0)}|/M_P$ 
for this choice of initial conditions, for several different values of $\phi^{(0)}$.
In the left panel, we normalize each $\phi(\tt)$ curve to the corresponding initial 
field value $\phi^{(0)}$ and adopt a logarithmic scale for the horizontal axis.
In the right panel, we show the same curves, but normalize each one to the fixed 
reference scale $M_P$ and adopt a linear scale for the horizontal axis. 

For $3H^{(0)}/2m \gg 1$, the Hubble-friction term in Eq.~\eqref{eq:EOM_reduced_2} 
is sufficiently large that $\phi$ is initially overdamped as it begins 
rolling from rest toward its potential minimum.  Within this ``slow-roll'' regime, 
$\phi''(\tt)$ is negligible in comparison to the other two terms in   
Eq.~\eqref{eq:EOM_reduced_2}, and therefore, to a good approximation, we have
\beq
  \phi' ~\approx~ - \frac{m}{3H}\phi\,.
\eeq
The solutions for $\phi$ and $H$ within the slow-roll regime
are therefore well approximated by 
\beqn
  \phi(\tt) &\approx& \phi^{(0)} 
    \exp\left[-\frac{1}{2}\int_{\tt^{(0)}}^{\tt}\frac{2m}{3H(\hatt)}\,d\hatt\, \right]
    \nonumber \\
  H(\tt) &\approx& H^{(0)} 
    \exp\left[-\frac{1}{2}\int_{\tt^{(0)}}^{\tt}\frac{2m}{3H(\hatt\,)}\,d\hatt\,\right]\,.
  \label{eq:H_slow}
\eeqn

Since $\phi$ evolves extremely slowly within this regime, $3H/(2m) \gg 1$ remains
large and $H(\tt) \approx H^{(0)}$ is effectively constant.  As a result, the 
universe experiences an epoch of accelerated expansion at early times.  This 
epoch effectively ends at the time $t_c$ at which $3H(t_c) = 2m$ 
and the coefficient of the Hubble-friction term in Eq.~\eqref{eq:EOM_reduced_2} 
drops below the value associated with critical damping.  At subsequent times 
$t > t_c$, the field experiences underdamped oscillations. 
The value of $\phi$ at $t_c$ is approximately independent of $\phi^{(0)}$ and
given by
\beq
  \phi(t_c)~\approx~ \frac{H(t_c)}{H^{(0)}}\,\phi^{(0)} 
    ~=~ \sqrt{\frac{8}{3}} M_P  \,,
  \label{eq:phi_c}
\eeq
as is evident from the right panel of Fig.~\ref{fig:phi_t_dom}.  However, as
is evident from the left panel, this implies that the extent to which the field 
is suppressed at $t_c$ relative to its initial value at $t^{(0)}$ becomes more 
severe as $\phi^{(0)}$ increases.  By contrast, in 
situations in which $3H^{(0)}/(2m) \ll 1$, such as that illustrated by the 
blue curve in each panel of the figure, the field is already underdamped at 
$t = t^{(0)}$, and oscillation commences immediately thereafter.

The difference between the initial time $t^{(0)}$ and the critical time $t_c$ can
be quantified in terms of the parameter $\Delta \tt_c\equiv \tt_c- \tt^{(0)}$, 
which can be estimated by evaluating Eq.~\eqref{eq:H_slow} at $\tt = \tt_c$.  Doing
this, we obtain
 \beq
  \Delta \tt_c ~\approx~ \expt{\frac{m}{3H}}_{t_c}^{-1} 
    \log\left(\frac{3H^{(0)}}{2m}\right)\,,
  \label{eq:Deltattc}
 \eeq
where $\expt{x}_{t_c}$ denotes the time-average of the quantity $x$ over the time interval
$\Delta t_c$.  Since $H$ decreases less rapidly than $t^{-1}$ as a function of time
while $\phi$ is slowly rolling, this time-average decreases with $H^{(0)}$.  It 
therefore follows from the form of Eq.~\eqref{eq:Deltattc} that $\Delta \tt_c$ occurs 
later for larger $H^{(0)}$.  The particular manner in which this delayed onset of 
oscillation manifests itself is illustrated in the right panel of Fig.~\ref{fig:phi_t_dom}.  
Indeed, by comparing the green, red and black curves shown in this panel, we observe that 
to a good approximation the functional forms of $\phi(\tt)$ obtained for different 
$\phi^{(0)}$ differ only by a horizontal shift.

\begin{figure}
\includegraphics[width=0.48\textwidth]{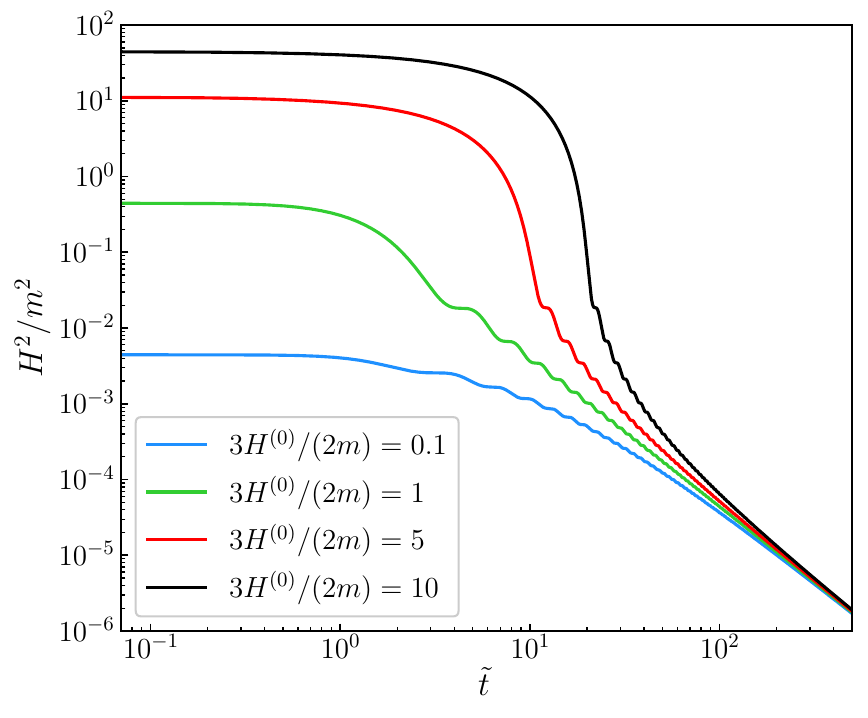}
\caption{The ratio $H^2/m^2$, which is proportional to the total energy density 
  $\rho_{\rm tot}$ of the universe, during an epoch wherein the energy density of 
  the universe is dominated by $\phi$ itself.  The different curves correspond to 
  the same parameter choices adopted in Fig.~\ref{fig:phi_t_dom}.  We observe that
  all of these curves --- despite the different choices of initial conditions they each
  represent --- asymptotically converge at late times to a power law which corresponds
  to a scaling behavior $\rho_{\rm tot} \sim a^{-3}$ at late times.
  \label{fig:H2m2_t_dom}}
\end{figure}

Eventually, at dimensionless times $\tt \gg \tt_c$, when $\phi(\tt)$ is deep within 
the oscillatory phase, the time-average of the equation-of-state parameter over 
a sufficiently long time window becomes $\expt{w_\phi}_t \approx 0$, much as it 
does in Case~I.  Thus, as in Case~I, we find $H\approx \kappa/(3t)$ with $\kappa=2$ for 
$\tt \gg \tt_c$.  However, since $H^2\propto \rho_\phi$ in Case~II, one finds that 
the total energy density of the universe approaches a universal functional 
form $\rho_{\rm \phi} \propto a^{-3}$ at late times, regardless of the choice of 
initial conditions.  This behavior is illustrated in Fig.~\ref{fig:H2m2_t_dom}, 
which shows the evolution of the dimensionless ratio $H^2/m^2$ for a variety of 
different choices of $3H^{(0)}/(2m)$.  
Indeed, we observe that all the curves shown in Fig.~\ref{fig:H2m2_t_dom} exhibit 
the same asymptotic functional form at large $\tt$.  

Looking forward, one of our primary concerns in this paper is to understand the 
manner in which the zero-modes of dynamically evolving scalar fields might contribute to 
the development of a stasis epoch within the cosmological timeline.  
It is clear from the above analysis that a single such scalar field --- whose zero-mode energy 
density transitions from slow-roll behavior to rapid oscillation over a relatively narrow 
time window --- cannot give rise to a stasis epoch alone.  However, as we shall see, many aspects
of the dynamics of individual scalars that we have highlighted in this section will play 
an important role in establishing and sustaining stasis 
when multiple such fields are considered.


\FloatBarrier
\section{Stasis from a tower of scalars\label{sec:ScalarTower}}


We shall now generalize the above analysis by replacing our single scalar field 
with an entire tower of scalar fields with different masses.  Our goal will be 
to determine whether an epoch of stasis might arise from such a tower and what 
its properties might be. 

\subsection{Preliminaries\label{sec:preliminaries}}

Let us now assume that there exists a tower of $N$ scalar fields $\phi_\ell$ in 
the early universe, each of which experiences a quadratic potential 
\beq
  V_\ell~=~ \frac{1}{2} m_\ell^2\phi_\ell^2\,,
  \label{eq:quadV}
\eeq
where the index $\ell=0, 1, 2,\dots, N-1$ labels the states in order of increasing 
mass.  The equation of motion for each state is then
\beq
  \ddot\phi_\ell+3H\dot\phi_\ell + m_\ell^2\phi_\ell~=~ 0~,
\label{eq:EOM_ell}
\eeq
while the energy density, pressure, equation-of-state parameter, and cosmological 
abundance of each state are given by 
\beqn
  \rho_\ell ~&=&~ \frac{1}{2}\dot\phi_\ell^2+\frac{1}{2}m_\ell^2\phi_\ell^2\nonumber\\
  P_\ell ~&=&~ \frac{1}{2}\dot\phi_\ell^2-\frac{1}{2}m_\ell^2\phi_\ell^2\nonumber\\
  w_\ell ~&\equiv &~ P_\ell/\rho_\ell\nonumber\\
  \Omega_\ell ~&\equiv&~ \frac{\rho_\ell}{3H^2M_P^2}~.
\label{eq:rho_P_w}
\eeqn
Each of these quantities is generally time-dependent.
We can also define the total abundance associated with our tower of states
 \beq
 \Omega_{\rm tow}(t)~\equiv~\sum_\ell 
       \Omega_\ell(t)~,
\eeq   
as well as the time-dependent effective equation-of-state parameter for our tower
\beq  
  \expt{w}(t)~\equiv~ 
    \frac{1}{\Omega_{\rm tow}(t)} 
    \, \sum_\ell \Omega_\ell(t)\,w_\ell(t)~.
\label{eq:w_eff_all}
\eeq
In general, we have $0\leq\Omega_{\rm tow}\leq 1$, with the value of $\Omega_{\rm tow}$ 
ultimately depending on what other energy components might also exist in the universe. 
Of course, if the total energy density of the universe is only that associated with the 
tower of $\phi_\ell$ states, we then have $\Omega_{\rm tow}=1$ at all times.  However, 
until stated otherwise, we shall not make this assumption. 

Along these lines, we note that $\Omega_{\rm tow}$ is not the only quantity whose 
value depends on the full energy content of the universe.   Indeed, even the individual 
abundances $\Omega_\ell$ implicitly depend on the full energy content through their  
dependence on $H$, or simply because abundances generally indicate the fraction of 
energy density {\it relative to the total energy density in the universe}\/.  Thus, 
for example, we see that the definition of $\langle w\rangle$ in Eq.~(\ref{eq:w_eff_all}) 
makes sense because it is invariant under such rescalings of the abundances. 

At any specific time $t$, certain states within the tower may still be overdamped while 
others may have already become underdamped.  We will respectively identify these groups of 
states as
\begin{itemize}
  \item {\it slow-roll}\/ components, which consist of the overdamped states with 
      $m_\ell<3H(t)/2$; and
  \item {\it oscillatory}\/ components, which are comprised of the underdamped states 
    with $m_\ell\geq 3H(t)/2$. 
\end{itemize}
Note that we shall use the terminology ``slow-roll'' (SR)  and ``oscillatory'' (osc) to 
indicate whether a given state is overdamped or underdamped regardless of whether its 
field VEV is actually rolling or oscillating.   Indeed, as we have seen in 
Sect.~\ref{sec:single_scalar}, a given state near the transition time may be underdamped 
and not yet have begun to oscillate;  likewise, a given state may be so severely overdamped 
that it is effectively stationary without any significant rolling behavior.

Let us define $\ell_c(t)$ to be the critical value of $\ell$ within the tower for which 
$3H(t)= 2m_\ell$.  More specifically, 
we shall implicitly assume that our spectrum of states is sufficiently dense that we 
may regard (or approximate) $\ell_c(t)$ as an integer at any time; this assumption will 
render our equations simpler but will not affect our final results.  We shall also 
consider the ``boundary'' state with $\ell=\ell_c(t)$ as just having become underdamped.
Thus, at any given time $t$, the states with $\ell<  \ell_c(t)$ are 
still overdamped, while those with $\ell\geq \ell_c(t)$ are underdamped. 
We then find that 
the total corresponding abundances at any time $t$ can be written as
\beqn 
  \Omega_{\rm SR}(t) ~&=&~ \sum_{\ell=0}^{\ell_c(t)-1}\Omega_\ell(t)\nonumber\\ 
  \Omega_{\rm osc}(t) ~&=&~ \sum_{\ell=\ell_c(t)}^{N-1}\Omega_\ell(t)~.
\label{Omegasums}
\eeqn
Likewise, we can define the total effective equation-of-state parameter associated with 
each of these separate groups of states: 
\beqn
 w_\SR(t)  \,&\equiv&\, 
 \frac{P_\SR(t)}{\rho_\SR(t)} \,=\,
   \frac{1}{\Omega_\SR(t) } 
   \sum_{\ell=0}^{\ell_c(t)-1}
   \Omega_\ell(t) \, w_\ell(t)\nonumber\\
  w_\osc(t)  \,&\equiv&\,
  \frac{P_\osc(t)}{\rho_\osc(t)}\,=\,
   \frac{1}{\Omega_\osc(t) } 
   \sum_{\ell_c(t)}^{N-1}
   \Omega_\ell(t) \, w_\ell(t)~.~~~~~~
\label{wSRosc_as_ell-sum}
\eeqn
Here $P_\SR$, $P_\osc$, $\rho_\SR$, and $\rho_\osc$ represent the total pressures 
and energy densities of each part of the tower, with the same summation limits 
as in Eqs.~(\ref{Omegasums}) and (\ref{wSRosc_as_ell-sum}).  It then follows that 
the effective equation-of-state parameter for the entire tower at any moment in 
time is given by
\beq
\langle w\rangle(t) ~=~ \frac{1}{\Omega_{\rm tow}(t)}\, 
  \biggl[ \Omega_\SR(t)\, w_\SR(t) 
  + \Omega_\osc(t) \, w_\osc(t)\biggr]~.
\label{langleranglew}
\eeq

Just as with a single scalar, the resulting dynamics of our system depends on whether 
we assume that the energy density of this entire scalar tower is subdominant to that of 
some other fixed energy component with a constant equation-of-state parameter. If so, 
then  the Hubble parameter evolves as $1/t$ and the results in 
Sect.~\ref{sec:fixed_background} can be directly applied here.  Each $\phi_\ell$ 
state will then simply evolve independently according to its own equation 
of motion Eq.~\eqref{eq:EOM_ell}, yielding solutions for the time-dependence of each $\phi_\ell$
which simply follow the analytical expressions in Eqs.~\eqref{eq:phi_bessel} and 
$\eqref{eq:phi_approx}$.  Indeed, all that is required is that we promote the 
coefficient $c_J$ and the dimensionless time variable $\tt$ to $\ell$-dependent 
quantities which essentially depend on the initial conditions and the mass spectrum of 
the $\phi_\ell$ states.

However, of far more interest is the situation in which the energy density of our tower of 
$\phi_\ell$ states is non-negligible, leading to a non-negligible value of $\Omega_{\rm tow}$.
In such circumstances, the effective equation-of-state parameter $\expt{w}$ of the entire 
tower will no longer generally be a time-independent quantity, since every state has a 
time-dependent equation-of-state parameter $w_\ell$.  Together with the unknown dynamics of 
the other energy components within the universe, it then follows that the Hubble parameter 
may not follow a simple $H\sim 1/t$ scaling relation.

The above situation would be greatly simplified if the universe were to evolve into an epoch 
of stasis during which the abundances of different cosmological energy components
remain constant despite cosmological expansion.
As a result, the effective equation-of-state parameter $w_{\rm univ}$ for the universe as 
a whole would then remain fixed.  This in turn implies that the Hubble parameter would indeed 
scale as $\sim 1/t$ during such an epoch.

However, there are many reasons to suspect that such a stasis epoch will no longer be possible.
In all of the previous studies of stasis~\cite{Dienes:2021woi,Dienes:2023ziv, Dienes:2022zgd},
the equation-of-state parameter associated with each energy component was treated as a 
constant. While appropriate for the situations under study in those works, in the present case we are 
dealing with a tower of fully dynamical $\phi_\ell$ fields.  Indeed, each of these individual 
fields has a complicated dynamics with its own time-dependent abundance $\Omega_\ell(t)$ 
and time-dependent equation-of-state parameter $w_\ell(t)$.  It is therefore not
{\it a priori}\/ clear whether these individual $w_\ell$-functions can conspire to produce 
a constant value of either $w_{\rm SR}$ or $w_{\rm osc}$.

\subsection{Parametrizing the scalar tower \label{subsec:tower_parameters}}

We shall shortly  determine an algebraic condition that must be satisfied in 
order for a stasis epoch to arise.   However, as we shall see, this condition 
will necessarily depend on the properties associated with our $\phi_\ell$ tower.

Different models of physics beyond the Standard Model (BSM) give rise to $\phi_\ell$ 
towers with different characteristic  properties.  In order to maintain generality and 
survey many models at once, we shall therefore adopt a useful 
parametrization~\cite{Dienes:2021woi,Dienes:2023ziv} which can simultaneously 
accommodate many different BSM scenarios.  In particular, the mass spectrum of the 
tower of states will be assumed to take the general form
\beq
  m_\ell~=~ m_0+ (\Delta m) \,\ell^\delta~,
  \label{eq:mass_spec}
\eeq
where $m_0$ is the mass of the lightest state and where $\Delta m$ and $\delta$ 
parametrize the mass splittings across the tower.  Such a mass spectrum is motivated 
by theories of extra spacetime dimensions, string theories, and strongly-coupled 
gauge theories.  For example, if the $\phi_\ell$ are the Kaluza-Klein (KK) excitations 
of a five-dimensional scalar in which one dimension of the spacetime is compactified 
on a circle of radius $R$ (or a $\mathbb{Z}_2$ orbifold thereof), we have either 
$\lbrace m_0,\Delta m,\delta\rbrace = \lbrace m, 1/R, 1\rbrace$ or 
$\lbrace m_0,\Delta m,\delta\rbrace =\lbrace m, 1/(2 m R^2), 2\rbrace$,
where $m$ denotes the four-dimensional scalar mass~\cite{Dienes:2011ja, Dienes:2011sa}.
This distinction depends on whether $m R \ll1$ or $mR\gg 1$, respectively.
Alternatively, if the $\phi_\ell$ are the bound states of a strongly-coupled gauge 
theory, we find that $\delta = 1/2$, where $\Delta m$ and $m_0$ are respectively 
determined by the Regge slope and intercept of the strongly-coupled 
theory~\cite{Dienes:2016vei}.  The same values also describe the excitations of a 
fundamental string.  Thus $\delta=\lbrace 1/2, 1,2\rbrace$ can serve as compelling 
``benchmark'' values.  

We shall likewise assume that the initial abundances of the $\phi_\ell$ states follow 
a power-law distribution
\beq
  \Omega^{(0)}_\ell~=~ \Omega^{(0)}_0\left(\frac{m_\ell}{m_0}\right)^\alpha\,,
  \label{eq:initial_abundance}
\eeq
where a superscript ``${(0)}$'' once again denotes the value of a quantity at the 
initial time $t=t^{(0)}$ and where $\Omega^{(0)}_0$ is the initial abundance of 
the lightest tower state.  Scaling relations of this form arise in a variety of
BSM scenarios which predict towers of states, and the exponent $\alpha$ in any
such scenario is ultimately determined by the mechanism through which the states 
in the tower are initially produced.  For example, production from the vacuum misalignment 
of a bulk scalar in a theory with extra spacetime dimensions predicts that 
$\alpha<0$~\cite{Dienes:2011ja, Dienes:2011sa}, while thermal freeze-out 
can accommodate either $\alpha>0$ or $\alpha<0$~\cite{Dienes:2017zjq}.  By 
contrast, if a tower of states is produced through the universal decay of a 
heavy particle, we have $\alpha=1$.  

Finally, since we are assuming that the energy density of each $\phi_\ell$ is 
dominated by the contribution from its spatially homogeneous zero-mode and that 
the contribution from particle-like excitations is negligible, each
$\rho_\ell^{(0)}$ is in general specified by the initial field value $\phi^{(0)}_\ell$ 
and its time derivative $\dot{\phi}^{(0)}_\ell$.  For simplicity --- and because the 
field velocities generated by many of the production mechanisms compatible with 
these assumptions are negligible --- we shall take $\dot{\phi}_\ell^{(0)}\approx 0$ 
for all $\ell$ in what follows. 

\subsection{Condition for stasis\label{subsec:stasis_condition}}

In order to determine the algebraic condition(s) under which stasis can emerge from a tower 
of dynamical scalars with these properties, we shall first posit --- as in previous
analyses~\cite{Dienes:2021woi,Dienes:2022zgd,Dienes:2023ziv} --- that the universe has 
indeed entered stasis.  We shall then assess the conditions under which this assumption is 
self-consistent, and finally indicate how our system actually evolves into the stasis state.

By definition, $\Omega_{\rm SR}$ and 
$\Omega_{\rm osc}$ must both remain effectively constant during stasis, as must the 
effective equation-of-state parameter for the universe as a whole.  This in turn 
implies that the Hubble parameter must also take the form $H \approx \kappa/(3t)$, 
where $\kappa$ is a constant, during stasis.  In what follows, 
we shall refer to the effectively constant stasis values for $\kappa$, $\Omega_{\rm SR}$, 
and $\Omega_{\rm osc}$  as $\barkappa$, $\barOmega_{\rm SR}$, and 
$\barOmega_{\rm osc}$, respectively.  We shall not make any assumptions
concerning the values of these stasis quantities, but rather determine how the 
self-consistency conditions for stasis constrain these values.

We begin by investigating the manner in which the various scalars within the tower 
are evolving at an arbitrary fiducial time $t_\ast \gg t^{(0)}$ by which the universe 
is already deeply in stasis and the Hubble parameter is evolving as 
$H=\barkappa/(3t)$.  We shall first focus on those $\phi_\ell$ fields which are still highly 
overdamped at $t = t_*$ and whose field values $\phi_\ell(t_\ast) \approx \phi_\ell^{(0)}$ 
are still approximately unchanged from their initial values.  The equation of 
motion for each such field is well approximated by Eq.~\eqref{eq:EOM_reduced},
and the solution to this equation therefore takes the same form as in 
Eq.~\eqref{eq:phi_approx}, but with a coefficient $c_\ell$ which depends 
on $\ell$:
\beq
  \phi_\ell(t)~ \approx~ c_{\ell} \, (m_\ell t)^{(1-\barkappa)/2}\,
    J_{(\barkappa-1)/2}(m_\ell t)~.
\eeq
Inserting these solutions into Eq.~\eqref{eq:rho_P_w} and using the Bessel-function 
recurrence relation 
\beq
  \frac{d}{dz}J_{\nu}(z)~=~ \frac{\nu}{z} J_{\nu}(z)-J_{\nu+1}(z)~,
\eeq
we obtain
\beqn
  \rho_\ell~&=&~\frac{1}{2}m_\ell^2c_\ell^2 (m_\ell t)^{1-\barkappa}
    \left[ J_{\frac{\barkappa+1}{2}}^2(m_\ell t) 
    + J_{\frac{\barkappa-1}{2}}^2(m_\ell t) \right]\, \nonumber\\
  P_\ell ~&=&~\frac{1}{2}m_\ell^2c_\ell^2 (m_\ell t)^{1-\barkappa}
    \left[ J_{\frac{\barkappa+1}{2}}^2(m_\ell t)
    - J_{\frac{\barkappa-1}{2}}^2(m_\ell t) \right]~.\nonumber\\
\label{eq:Prho_ell}
\eeqn

Our assumption that the initial abundances of the $\phi_\ell$ scale with $m_\ell$
according to Eq.~\eqref{eq:initial_abundance} specifies the corresponding
scaling relation for the $c_\ell$.  Since any $\phi_\ell$ which is highly overdamped 
at $t = t_\ast$ is even more highly overdamped at $t=t^{(0)}$, it follows that 
$m_\ell t^{(0)} \ll 1$ for such a field.  The initial energy density $\rho_\ell^{(0)}$ 
of any such field is therefore well approximated by the $m_\ell t\rightarrow 0$ limit 
of the expression for $\rho_\ell$ in Eq.~\eqref{eq:Prho_ell} with $m_\ell$ held fixed.
We thus have
\begin{equation}
  \rho_\ell^{(0)} ~\approx~ \lim_{m_\ell t\to 0} \rho_\ell ~=~ 
    \frac{1}{2}c_\ell^2m_\ell^2\mathcal{J}(\barkappa)~,
  \label{eq:Lim_rho_SR_init}
\end{equation}
where the quantity $\mathcal{J}(\barkappa)$ is independent of $\ell$ and given by 
\begin{equation}
  \mathcal{J}(\barkappa) ~\equiv
    ~\frac{2^{1-\barkappa}}{\Gamma^2\left(\frac{\barkappa+1}{2} \right)}~,
\end{equation}
where $\Gamma(z)$ denotes the Euler gamma function.  Comparing 
the form of $\rho_\ell^{(0)}$ in Eq.~(\ref{eq:Lim_rho_SR_init}) with the expression 
for $\Omega_\ell$ in Eq.~\eqref{eq:initial_abundance}, we find that
\begin{equation}
    c_\ell^2 ~=~ \frac{2\rho_0^{(0)}}{m_\ell^2\mathcal{J}(\barkappa)}
      \left(\frac{m_\ell}{m_0}\right)^\alpha~
\end{equation}
and therefore that
\begin{eqnarray}
  \rho_\ell &\approx& \frac{\rho_0^{(0)}}{\mathcal{J}(\barkappa)}
    \left(\frac{m_\ell}{m_0}\right)^{\alpha} (m_\ell t)^{1-\barkappa} 
    \nonumber \\ & & ~~~~\times
    \left[ J_{\frac{\barkappa+1}{2}}^2(m_\ell t) 
    + J_{\frac{\barkappa-1}{2}}^2(m_\ell t) \right]\,.
\label{eq:rho_ell_Bessel_alpha}  
\end{eqnarray}

The total energy density associated with the slow-roll states at a given time $t$ is 
simply the sum of the contributions from the individual states which still remain  
overdamped at that time:
\beq
  \rho_{\rm SR}~=~\sum_{\ell=0}^{\ell_c(t)-1}\rho_\ell\,.
  \label{eq:rho_SR_raw}
\eeq
Within the regime in which the density of states per unit mass is large --- and the difference 
between the times at which each pair of adjacent states $\phi_{\ell}$ and $\phi_{\ell-1}$ 
undergo their critical-damping transitions is therefore small --- we may obtain a reliable 
approximation for $\rho_{\rm SR}$ by working in the continuum limit in which the discrete 
index $\ell$ is promoted to a continuous variable and the sum in Eq.~(\ref{eq:rho_SR_raw}) 
becomes an integral.  In particular, as discussed in more detail in 
Refs.~\cite{Dienes:2021woi,Dienes:2023ziv}, this limit corresponds to simultaneously taking 
\beq
  \Delta m\to 0\,,~~~N\to \infty
\label{continuumlimit}
\eeq
and
\beq
  m_0\to 0\,,~~~m_{N-1}\to \infty
\label{endpointslimits}
\eeq
while holding the ratio $\Delta m/m_0$ fixed.  In this limit, the sum in 
Eq.~(\ref{eq:rho_SR_raw}) becomes an integral over the continuous variable $\ell$ --- or, 
equivalently, over the continuous mass variable $m$ obtained from this $\ell$ via 
Eq.~(\ref{eq:mass_spec}) --- and $\rho_{\rm SR}$ takes the form
\beqn
  \rho_{\rm SR} &=&
    \frac{\rho_0^{(0)}I^{(\rho)}_{\rm SR}(\barkappa)}
      {\delta\Delta m^{1/\delta}m_0^\alpha\mathcal{J}(\barkappa)}\,
    \frac{1}{t^{\alpha+1/\delta}}\,,
  \label{eq:rho_bessel_integral}
\eeqn
where we have defined
\begin{eqnarray}
  I^{(\rho)}_{\rm SR}(\barkappa) &\equiv&
      \int_0^{m_{\ell_c}}dm ~ t (mt)^{\alpha+1/\delta-\barkappa} \nonumber \\
    & & ~~~~~~ \times \left[ J_{\frac{\barkappa+1}{2}}^2(mt) 
      + J_{\frac{\barkappa-1}{2}}^2(mt) \right]\,.~~~
  \label{eq:bessel_int_rho_V_prelim}
\end{eqnarray}
Changing integration variables from $m$ to $\tt \equiv mt$ and noting that
the upper limit of integration in the resulting integral can be expressed as
as $m_{\ell_c} t = 3Ht/2 = \barkappa/2$ during stasis, we find that 
$I^{(\rho)}_{\rm SR}(\barkappa)$ is in fact independent of $t$ and given by
\begin{equation}    
  I^{(\rho)}_{\rm SR}(\barkappa) ~=~ 
    \int_0^{\barkappa/2}d\tt~\tt^{\alpha+1/\delta-\barkappa}
    \left[ J_{\frac{\barkappa+1}{2}}^2(\tt) 
    + J_{\frac{\barkappa-1}{2}}^2(\tt) \right] \,.
  \label{eq:bessel_integral_rho_V}
\end{equation}

Since the abundance $\Omega_{\rm SR}$ of the slow-roll component must by definition 
remain constant while the universe is in stasis, the expression for $\rho_{\rm SR}$ 
in Eq.~(\ref{eq:rho_bessel_integral}) implies a consistency condition on the values of 
the scaling exponents $\alpha$ and $\delta$.  Indeed, since $H = \barkappa/(3t)$ 
during stasis, this abundance is given by
\begin{equation}
  \Omega_{\rm SR} \,=\, \frac{\rho_{\rm SR}}{3M_P^2H^2} 
    \,=\, \frac{3\rho_0^{(0)}I^{(\rho)}_{\rm SR}(\barkappa)}
    {\barkappa^2 \delta\Delta m^{1/\delta}m_0^\alpha M_P^2\mathcal{J}(\barkappa)}
    \frac{1}{t^{\alpha+1/\delta-2}}\, .~
\label{eq:OmegaBarSR_for_scaling}
\end{equation}
We thus find that in order for $\Omega_{\rm SR}$ to be independent of $t$ during stasis, 
our scaling exponents $\alpha$ and $\delta$ must satisfy
\beq
 \alpha+\frac{1}{\delta}~=~2~.
  \label{eq:stasis_constraint}
\eeq
Indeed, since $\rho^{(0)}_\ell\sim m_\ell^\alpha$, this stasis condition
implies that our initial field displacements  for $\ell\gg 1$
must exhibit the universal $\delta$-independent behavior
\beq 
 \phi^{(0)}_\ell~\sim~ \ell^{-1/2}~,
 \label{displacementscaling}
\eeq
with increasingly small initial field displacements as one proceeds up the tower. 
Thus, we see that even for $\alpha>0$, stasis never requires growing initial field 
displacements.

An expression for the 
pressure $P_{\rm SR}$ associated with the slow-roll component may be obtained through a 
procedure analogous to that which we used in obtaining our expression for $\rho_{\rm SR}$ 
in Eq.~(\ref{eq:rho_bessel_integral}).  In particular, one finds that
\beqn
  P_{\rm SR}
    ~=~\frac{\rho_0^{(0)}I_{\rm SR}^{(P)}(\barkappa)}
      {\delta\Delta m^{1/\delta}m_0^\alpha\mathcal{J}(\barkappa)}\,
    \frac{1}{t^{\alpha+1/\delta}}\,,
  \label{eq:bessel_integral_P_V}
\eeqn
where we have defined
\beq
  I_{\rm SR}^{(P)}(\barkappa) ~\equiv~ 
    \int_0^{\barkappa/2}d\tt~\tt^{\alpha+1/\delta-\barkappa}
    \left[ J_{\frac{\barkappa+1}{2}}^2(\tt)-J_{\frac{\barkappa-1}{2}}^2(\tt) \right]\,.
  \label{eq:Int_SR_P}
\eeq
It therefore follows that the value $\overline{w}_{\rm SR}$ of the equation-of-state 
parameter for the slow-roll component during stasis is indeed time-independent and given by 
\beq
  \overline{w}_{\rm SR}~=~\frac{I_{\rm SR}^{(P)}(\barkappa)}
    {I_{\rm SR}^{(\rho)}(\barkappa)}~.
  \label{eq:w_sr}
\eeq

We now turn to consider the fields which are underdamped at $t = t_*$.  In general, 
the heavier such fields could have been either underdamped or overdamped at $t = t^{(0)}$,
depending on the relationship between $m_{N-1}$ and $H^{(0)}$.
However, since the energy density of an individual $\phi_\ell$ which is underdamped
during any particular time interval decreases over time relative to that of any state 
which is overdamped during that interval, the collective contribution to $\rho_{\rm osc}$ 
from those $\phi_\ell$ which are already underdamped at $t^{(0)}$ decreases over time 
and eventually becomes negligible in comparison to the collective contribution from the 
$\phi_\ell$ which begin oscillating after $t^{(0)}$.  

Given this observation, we shall take 
our fiducial time $t_\ast$ to be sufficiently late that $\rho_{\rm osc}$ is dominated at
this time by the collective contribution from those $\phi_\ell$ states which were not only 
overdamped at $t = t^{(0)}$ but also still overdamped at the time the stasis epoch began.  
Since these $\phi_\ell$ began oscillating only after the Hubble parameter was effectively 
given by $H \approx \barkappa/(3t)$, their individual energy densities are well 
approximated by Eq.~(\ref{eq:rho_ell_Bessel_alpha}).  Moreover, since the collective 
contribution to $\rho_{\rm osc}$ from fields which began oscillating before the stasis 
epoch began is negligible at $t = t_\ast$, we may approximate $\rho_{\rm osc}$ at
this time --- or indeed at any time $t$ at which the universe is likewise sufficiently 
deeply in stasis that these conditions are satisfied --- by the sum  
\begin{equation}
  \rho_{\rm osc}~\approx~\sum_{\ell=\ell_c(t)}^{N-1}\rho_\ell\,.
  \label{eq:rho_osc_raw}   
\end{equation}
An expression for the pressure $P_{\rm osc}$ may be derived in an analogous manner.     
In the continuum limit, these expressions evaluate to
\beqn
  \rho_{\rm osc}&=&\frac{\rho_0^{(0)}I_{\rm osc}^{(\rho)}(\barkappa)}
    {\delta\Delta m^{1/\delta}m_0^\alpha\mathcal{J}(\barkappa) }
    \frac{1}{t^{\alpha+1/\delta}}\nonumber\\
  P_{\rm osc}&=&\frac{\rho_0^{(0)}I_{\rm osc}^{(P)}(\barkappa)}
    {\delta\Delta m^{1/\delta}m_0^\alpha\mathcal{J}(\barkappa) }
    \frac{1}{t^{\alpha+1/\delta}}\,,
  \label{eq:rho_osc_and_P_osc}
\eeqn
where $I_{\rm osc}^{(\rho)}$ and $I_{\rm osc}^{(P)}$ are expressions of exactly
the same form as the expressions for $I_{\rm SR}^{(\rho)}$ and $I_{\rm SR}^{(P)}$ in 
Eqs.~(\ref{eq:bessel_integral_rho_V}) and~(\ref{eq:bessel_integral_P_V}), respectively, 
but with the lower limit of integration replaced by $\barkappa/2$ and the upper limit of 
integration replaced by $m_{N-1}t \to \infty$ in each case.  For all $\barkappa\geq 2$,
the integrals in these expressions converge.  

The form of $\rho_{\rm osc}$ in Eq.~(\ref{eq:rho_osc_and_P_osc}) implies that for values 
of $\alpha$ and $\delta$ which satisfy the condition in Eq.~(\ref{eq:stasis_constraint}),
the corresponding abundance $\Omega_{\rm osc}$ is constant during stasis.  It also 
follows from Eq.~(\ref{eq:rho_osc_and_P_osc}) that the effective equation-of-state 
parameter $w_{\rm osc}$ remains effectively constant during stasis at the value 
$\overline{w}_{\rm osc}$, where
\beq
  \overline{w}_{\rm osc}~=~
    \frac{I_{\rm osc}^{(P)}(\barkappa)}{I_{\rm osc}^{(\rho)}(\barkappa)}\,.
  \label{eq:w_osc}
\eeq

The effective equation-of-state parameter $\langle w\rangle$ for the tower as a whole is also
essentially constant during stasis.  Indeed, this constant value, which we denote $\barw$,  
can be obtained from Eq.~(\ref{langleranglew}) by taking $\Omega_{\rm SR}$, $\Omega_{\rm osc}$,
and the equation-of-state parameters $w_{\rm SR}$ and $w_{\rm osc}$ equal to their stasis 
values.  In particular, we find that
\beq
  \barw ~=~ \frac{I_{\rm osc}^{(P)}(\barkappa)+I_{{\rm SR}}^{(P)}(\barkappa)}
    {I_{\rm osc}^{(\rho)}(\barkappa)+I_{\rm SR}^{(\rho)}(\barkappa)}\,. 
\label{first_time2}
\eeq 
  
\begin{figure}
\includegraphics[width=0.48\textwidth]{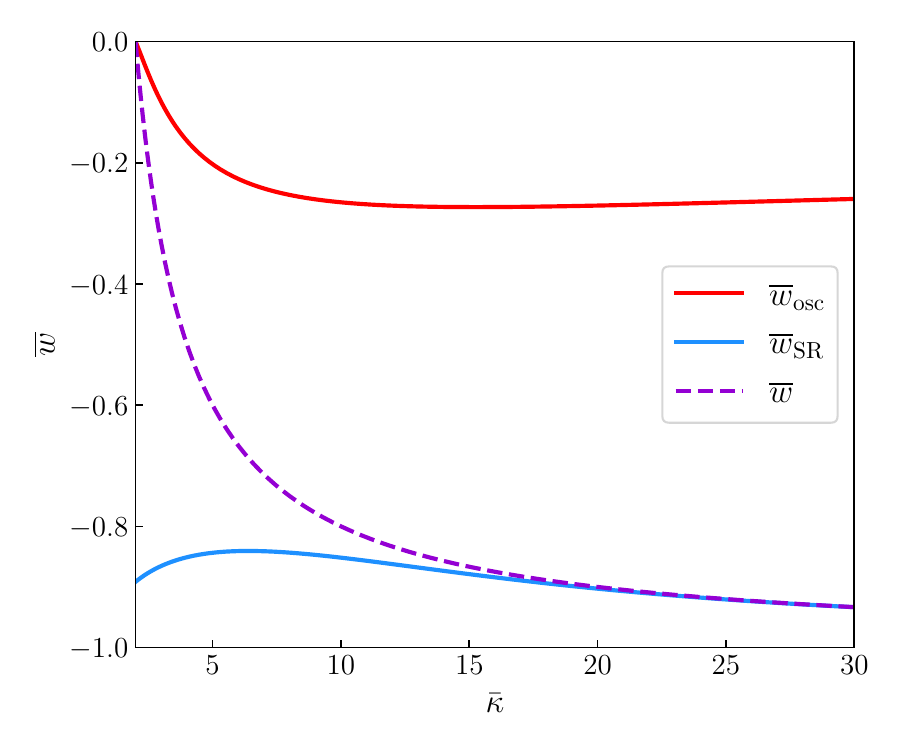}
\caption{
The stasis equation-of-state parameters $\barw_{\rm SR}$ and $\barw_{\rm osc}$ for 
our slow-roll and oscillatory energy components, along with the equation-of-state 
parameter $\barw$ for the tower of scalar fields as a whole, plotted as functions of 
the parameter $\barkappa$ within the range $2 \leq \barkappa \leq 30$.}
\label{fig:w_kappa}
\end{figure}

In Fig.~\ref{fig:w_kappa}, we show how this effective equation-of-state parameter 
$\barw$, along with the equation-of-state parameters $\barw_{\rm SR}$ and 
$\barw_{\rm osc}$ for the slow-roll and oscillatory components, vary as 
functions of $\barkappa$ within the range $2 \leq \barkappa \leq 30$.
Perhaps most notably, these results reveal the extent to which the effective 
equation-of-state parameters $\barw_{\rm SR}$ and $\barw_{\rm osc}$ differ from the 
characteristic values associated with vacuum energy ($w=-1$) and for matter ($w=0$), 
respectively, across nearly the entire range of $\barkappa$ shown.  The difference
between $\barw_{\rm SR}$ and the equation-of-state parameter for vacuum energy owes
primarily to the fact that $w_{\rm SR}$ includes contributions from fields which, 
while still slowly rolling, nevertheless have non-negligible field velocities 
$\dot{\phi}_\ell$ and therefore also have $w_\ell > -1$.  The difference between 
$\barw_{\rm osc}$ and the equation-of-state parameter for matter owes to the fact 
that $\barw_{\rm osc}$ includes contributions not only from heavier $\phi_\ell$ 
which are already oscillating rapidly and whose equation-of-state parameters are 
therefore also varying rapidly within the range $-1 \leq \ w_\ell \leq 1$, but also 
from lighter $\phi_\ell$ which have only recently transitioned from overdamped to 
underdamped evolution.  While the former contributions sum incoherently to zero, 
the latter contributions in general do not. 
Moreover, since the contribution that each $\phi_\ell$ makes to 
$\barw_{\rm osc} = \sum_{\ell=\ell_c}^{N-1}\Omega_\ell w_\ell$ is weighted by its 
abundance, the contributions from the $\phi_\ell$ which have only recently transitioned 
from overdamped to underdamped evolution and thus still have negative values of 
$w_\ell$ have a greater impact on this effective equation-of-state parameter.  As a 
result, $\barw_{\rm osc} < 0$ for all $\barkappa > 2$.  

We also observe from Fig.~\ref{fig:w_kappa} that
the effective equation-of-state parameter for the tower as a whole interpolates 
between $\barw_{\rm SR}$ and $\barw_{\rm osc}$, with $\barw$ approaching 
$\barw_{\rm osc}$ as $\barkappa\rightarrow 2$ and approaching 
$\barw_{\rm SR}$ as $\barkappa\rightarrow\infty$.  As $\barkappa\rightarrow 2$,
we see that $\barw \rightarrow 0$ and the tower behaves effectively like massive
matter.  By contrast, as $\barkappa\rightarrow \infty$, we find that 
$\barw \rightarrow - 1$ and the tower behaves like vacuum energy.

The abundance $\barOmega_{\rm SR}$ of the slow-roll component during stasis can be
obtained by applying the constraint in Eq.~(\ref{eq:stasis_constraint}) to the expression
in Eq.~(\ref{eq:OmegaBarSR_for_scaling}).  Noting that 
$\rho_0^{(0)} = m_0^2(\phi_0^{(0)})^2/2$, we may express this abundance as
\beq
  \barOmega_{\rm SR}~=~\frac{\rho_{\rm SR}}{3M_P^2H^2}~=~\frac{3}{2}
    \frac{I_{\rm SR}^{(\rho)}(\barkappa)}{\delta\barkappa^2\mathcal{J}(\barkappa)}
    \left(\frac{m_0}{\Delta m}\right)^{1/\delta}\left(\frac{\phi_0^{(0)}}{M_P}\right)^2\,.
  \label{eq:Omega_sr_1}
\eeq
Thus, we find that $\barOmega_{\rm SR}$ exhibits an explicit dependence on the 
initial value of the lightest field in the tower.

Alternatively, $\barOmega_{\rm SR}$ may also be expressed in terms of the ratio 
$H^{(0)}/m_{N-1}$ of the initial value of the Hubble parameter to the mass of the heaviest
scalar in the tower --- a ratio which carries information the extent to which 
this scalar is damped at $t = t^{(0)}$.  Indeed, in the continuum limit, one finds that 
the total initial abundance of the tower is given by
\beq
  \Omega_{\rm tow}^{(0)} ~=~ \frac{m_{N-1}^2}{2\delta m_0^2}
    \left(\frac{m_0}{\Delta m}\right)^{1/\delta}\Omega_0^{(0)}\,
\eeq
and that the initial energy density $\rho_0^{(0)} = 3M_P^2(H^{(0)})^2\Omega_0^{(0)}$ 
of the lightest state in the tower may therefore be expressed as
\beq
  \rho_0^{(0)} ~=~ 6\delta\left(\frac{M_P H^{(0)} m_0}{m_{N-1}}\right)^2
    \left(\frac{\Delta m}{m_0}\right)^{1/\delta}
    \Omega_{\rm tow}^{(0)}\,.
\eeq
Substituting this expression for $\rho_0^{(0)}$ into Eq.~(\ref{eq:OmegaBarSR_for_scaling}) 
and applying the constraint in Eq.~(\ref{eq:stasis_constraint}), we obtain 
\beq
  \barOmega_{\rm SR}~=~
    \frac{18I_{\rm SR}^{(\rho)}(\barkappa)}{\barkappa^2\mathcal{J}(\barkappa)}
    \left(\frac{H^{(0)}}{m_{N-1}}\right)^2\Omega_{\rm tow}^{(0)}\,.
  \label{eq:Omegabar_condition}
\eeq
Likewise, we note that the stasis abundance for the oscillatory component is 
given by
\beq
  \barOmega_{\rm osc}~=~
    \frac{18I_{\rm osc}^{(\rho)}(\barkappa)}{\barkappa^2\mathcal{J}(\barkappa)}
    \left(\frac{H^{(0)}}{m_{N-1}}\right)^2\Omega_{\rm tow}^{(0)}\,.
  \label{eq:Omegabar_osc_codecil}
\eeq

Taken together, Eqs.~(\ref{eq:Omega_sr_1}) and~(\ref{eq:Omegabar_condition}) imply that
\beq
  \frac{H^{(0)}}{m_{N-1}}\sqrt{\Omega^{(0)}_{\rm tow}} ~=~ \sqrt{\frac{1}{12\delta}} 
  \left(\frac{m_0}{\Delta m}\right)^{1/(2\delta)}\frac{\phi_0^{(0)}}{M_P}\,,
  \label{eq:HOvermN_to_phiOverMP}
\eeq
for any value of $m_{N-1}$.  Indeed, straightforward calculation confirms that this relation 
holds even in the $m_{N-1}\to \infty$ limit, and that $\barOmega_{\rm SR}$ therefore remains 
finite in this limit as well.  For a given aggregate initial abundance 
$\barOmega_{\rm tow}^{(0)}$ for the tower, then, we may treat $\barOmega_{\rm SR}$ as a function 
of two dimensionless parameters: $\barkappa$ and either $H^{(0)}/m_{N-1}$ or $\phi_0^{(0)}/M_P$. 

Thus, to summarize the results of this section, 
we have shown that as long as the condition in Eq.~(\ref{eq:stasis_constraint}) is 
satisfied, the system of dynamical equations which govern the evolution of our $\phi_\ell$ in 
the early universe permits a stasis solution wherein the aggregate abundances $\Omega_{\rm SR}$ 
and $\Omega_{\rm osc}$ both remain effectively constant.  Somewhat miraculously, such a stasis 
solution emerges despite the fact that the equation-of-state parameters $w_\ell$ for the individual 
$\phi_\ell$ evolve non-trivially with time as each such field transitions dynamically from the
overdamped to the underdamped phase.  Of course, many approximations were made on the road 
to the result in Eq.~(\ref{eq:stasis_constraint}).  These include, for example, the transition 
to a continuum limit in Eq.~(\ref{continuumlimit}) and the subsequent approximations for the 
summation endpoints in Eq.~(\ref{endpointslimits}).
However, it turns out that none of these approximations affect the manner in which our
expression for $\Omega_\SR$ in Eq.~(\ref{eq:OmegaBarSR_for_scaling}) and the 
analogous expression for $\Omega_\osc$ depend on $t$.  As a result, the constraint which 
cancels this time dependence --- namely that in Eq.~(\ref{eq:stasis_constraint}) ---
is an exact constraint that does not require any modification.
Indeed, these approximations only affect the {\it prefactors}\/ that are associated 
with these expressions, and we shall see that these prefactors are of lesser concern 
because changes to their precise values disturb neither the existence of the stasis state nor 
the ability of the universe to evolve into it.   These issues are discussed more 
fully in Ref.~\cite{Dienes:2023ziv}.

It is interesting to compare the stasis condition for this system to the 
stasis condition derived in Ref.~\cite{Dienes:2023ziv} for an analogous system 
consisting of a tower of $\phi_\ell$ states undergoing underdamping transitions 
in which each of the $\phi_\ell$ was modeled as having a {\it fixed}\/ common 
equation-of-state parameter $w_\ell = w$ prior to the instant at which the 
critical-damping transition occurs and as having a {\it fixed}\/ equation-of-state parameter 
$w_\ell=0$ thereafter.  Taking the $w\to -1$ limit in this system would then correspond to 
treating each $\phi_\ell$ as pure vacuum energy prior to its underdamping transition 
and to treating each $\phi_\ell$ after as pure matter afterwards.  For general values 
of $w$ lying within the range $-1< w<0$, it was then found that the condition for 
stasis is~\cite{Dienes:2023ziv}
\beq
   \alpha + \frac{1}{\delta} ~=~ 2- (1+ w)\barkappa~.
\label{fixedwresult}
\eeq
Comparing this result with that in Eq.~(\ref{eq:stasis_constraint}), we see that these two 
stasis conditions coincide precisely when $w= -1$.  This then lends credence to both 
approaches to studying such towers of dynamical scalars and indicates that they are mutually 
consistent in the $w\to -1$ limit, which corresponds to a true vacuum-energy/matter stasis.

That said, there is an important fundamental difference between the results in 
Eqs.~(\ref{eq:stasis_constraint}) and (\ref{fixedwresult}): given an input value for 
$\alpha+1/\delta$, the former constraint does not predict a particular stasis value for 
$\barkappa$ (or equivalently for $\overline{w}$), 
while the latter does.  Or, phrased somewhat differently, our derivation of the stasis condition 
in Eq.~(\ref{eq:stasis_constraint}) made absolutely no assumption concerning what {\it other}\/ energy 
components might also simultaneously exist in the universe, so long as the entire 
universe experiences a net stasis with the Hubble parameter taking the form 
$H(t)\approx \barkappa/(3t)$ for some constant $\barkappa$.  In particular, 
it was not necessary to impose any relationship between the value of $\barkappa$ 
and the abundances $\barOmega_{\rm SR}$ and $\barOmega_{\rm osc}$ --- or equivalently 
between the total energy density $\rho_{\rm tot}$ of the universe and the contributions 
to $\rho_{\rm tot}$ which come from the tower states $\phi_\ell$ alone.  We shall find in 
Sect.~\ref{sec:tracker} that this fundamental difference has profound consequences.

\subsection{Stasis in a tower-dominated universe}\label{sec:tower_alone}

Throughout this section, we have utilized the fact that the Hubble parameter 
during stasis takes the general form $H=\barkappa/(3t)$.  However, up to this 
point, we have made no assumptions concerning the value of $\barkappa$.  
In general, $\barkappa$ is directly related to the stasis value 
$\barw_{\rm univ}$ of the equation-of-state parameter $w_{\rm univ}$ for the universe 
as a whole.  In order to show this, we begin by noting that we can implicitly define a 
time-dependent parameter $\kappa(t)$ via the relation 
\beq
  \frac{dH}{dt} ~=~ -\frac{3}{\kappa}  H^2 ~.
  \label{eq:Def_of_kappa_of_t}
\eeq
Indeed,  during a stasis epoch in which $\kappa(t)$ is effectively constant 
with a value $\barkappa$,  we recover from Eq.~(\ref{eq:Def_of_kappa_of_t})  
the stasis relation $H \approx \barkappa/(3t)$.  However, the Friedmann 
acceleration equation for a flat universe generally tells us 
\beq
  \frac{dH}{dt} ~=~ -\frac{3}{2} H^2 \left(1 + {w}_{\rm univ}\right)~.
\label{eq:comparee} 
\eeq
Comparing Eqs.~(\ref{eq:Def_of_kappa_of_t}) and (\ref{eq:comparee}) then yields 
the general relation $\kappa = 2/(1+w_{\rm univ})$, where $\kappa$ and 
$w_{\rm univ}$ are in general both time-dependent quantities.  During stasis,
both of these quantities are effectively constant and we therefore have 
\beq
  \barkappa ~=~ \frac{2}{1+\barw_{\rm univ}}\,.
\label{eq:kappa_univ}
\eeq

While Eq.~(\ref{eq:kappa_univ}) provides a general relation between 
$\barkappa$ and $\barw_{\rm univ}$, both of which describe the universe as a whole, 
we have not yet asserted any relation between 
these quantities and quantities such as $\overline{w}$ 
or $\barOmega_{\rm tow}$ which describe the tower itself.  In other words, 
we have made no assumption about whether our scalar tower constitutes the entirety 
of the energy density of the universe, or whether there exist additional energy 
components during stasis as well.  Such an assumption --- and additional details 
concerning the abundances and equation-of-state parameters of any such energy 
components --- would be necessary before any such relation between $\barkappa$ 
and $\overline{w}_{\rm univ}$ on the one hand, and quantities such as 
$\overline{w}$ and $\barOmega_{\rm tow}$ on the other hand, could be formulated.
Thus, in order to proceed further, we must specify whether any additional 
energy components are present during stasis and what their properties might be.

For the remainder of this section, we shall focus on the simplest case --- that 
in which the tower states collectively represent the entirety of the energy density 
in the universe.  In other words, we shall assume that  $\Omega_{\rm tow}(t)=1$ for 
all $t$ and defer our study of the more general case in which additional cosmological 
energy components are present to Sect.~\ref{sec:tracker}.

In the absence of additional energy components, we have 
$\barOmega_{\rm osc} = 1-\barOmega_\SR$.  Likewise, the equation-of-state parameter 
for the universe as a whole during stasis is simply $\barw_{\rm univ} = \barw$, 
where in this case  
\beq
  \barw ~=~ \barw_{\rm SR}\barOmega_{\rm SR} +\barw_{\rm osc}\barOmega_{\rm osc}\,.
  \label{first_time_again}
\eeq
It therefore follows from Eq.~\eqref{eq:kappa_univ} that 
\beq 
  \barkappa ~=~ \frac{2}{1+\barw_{\rm osc}(\barkappa)
    +[\barw_{\rm SR}(\barkappa) -\barw_{\rm osc}(\barkappa)] 
    \barOmega_{\rm SR}}\,,
  \label{eq:kappa_Omega_1} 
\eeq 
where we have included the explicit dependence of $\barw_{\rm SR}$ and 
$\barw_{\rm osc}$ on $\barkappa$ in this expression in order to emphasize that 
these quantities not only depend on, but are indeed completely specified by, 
the value of $\barkappa$.  Substituting the expression for $\barOmega_{\rm SR}$ 
in Eq.~(\ref{eq:Omegabar_condition}) with $\Omega^{(0)}_{\rm tow}=1$ into this 
equation, we arrive at a transcendental equation for $\barkappa$.  This
equation, which takes the form
\beq
  \frac{2\barkappa-[1+\barw_{\rm osc}(\barkappa)]\barkappa^2}
    {\barw_{\rm SR}(\barkappa)-\barw_{\rm osc}(\barkappa)}
    \,=\,\frac{18I_{\rm SR}^{(\rho)}(\barkappa)}{\mathcal{J}(\barkappa)}
      \left(\frac{H^{(0)}}{m_{N-1}}\right)^2\,,
\label{eq:solve_kappa_1}
\eeq
may be solved numerically for any given value of the ratio $H^{(0)}/m_{N-1}$.
From this solution, the corresponding values of $\barOmega_{\rm SR}$, 
$\barOmega_{\rm osc}$, $\barw_{\rm SR}$, and $\barw_{\rm osc}$ may be obtained
in a straightforward manner.

\begin{figure}
\centering
\includegraphics[width=0.48\textwidth]{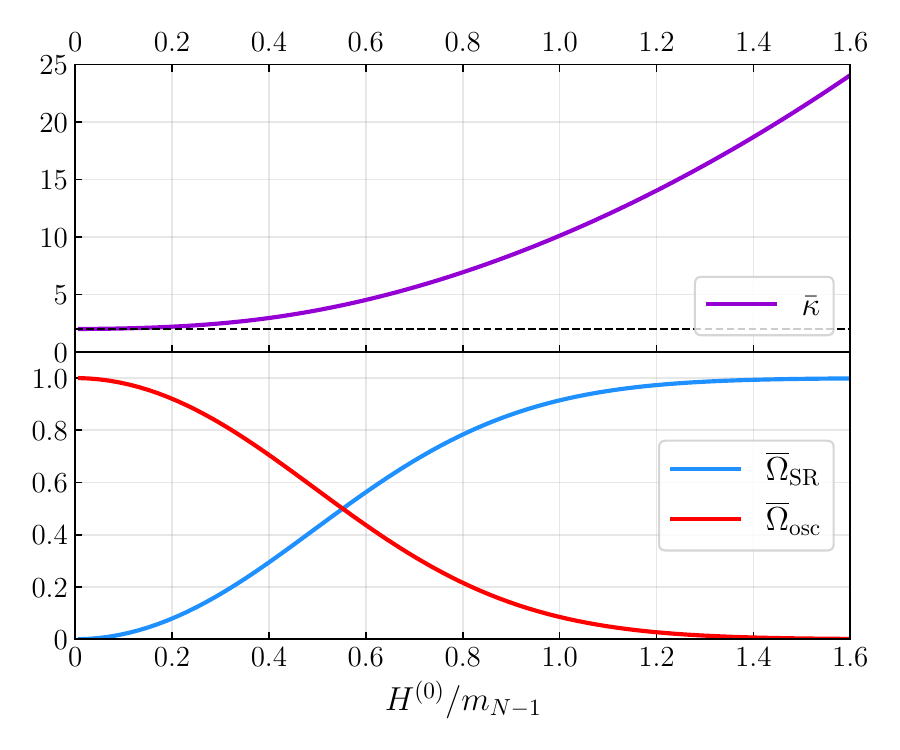}
\caption{The value of $\barkappa$ (upper panel) and the values of $\barOmega_\SR$ 
  and $\barOmega_\osc$ (lower panel), plotted as functions of the ratio 
  $H^{(0)}/m_{N-1}$ for the case in which no additional energy components are 
  present during stasis.
  \label{fig:Omega_ratio}}
\end{figure}

In Fig.~\ref{fig:Omega_ratio}, we show both the value of $\barkappa$ (upper panel)
and the values of $\barOmega_{\rm SR}$ and $\barOmega_{\rm osc}$ (lower panel) as
functions of $H^{(0)}/m_{N-1}$.  We observe from the upper panel that $\barkappa$ 
approaches the value $\barkappa = 2$ associated with a matter-dominated universe 
in the $H^{(0)}/m_{N-1}\rightarrow 0$ limit, but grows without bound as 
$H^{(0)}/m_{N-1}$ increases.  Accordingly, we observe from the lower panel that 
$\barOmega_{\rm SR}\rightarrow 0$ in the $H^{(0)}/m_{N-1}\rightarrow 0$ limit.  However,
this abundance increases monotonically with $H^{(0)}/m_{N-1}$ and approaches 
unity as $H^{(0)}/m_{N-1}\rightarrow \infty$.  

\subsection{Dynamical evolution and attractor behavior}

Having established the conditions under which stasis can emerge in our 
dynamical-scalar scenario, and having determined how the stasis abundances
$\barOmega_{\rm SR}$ and $\barOmega_{\rm osc}$ depend on input parameters, we 
now examine whether $\Omega_{\rm SR}$ and $\Omega_{\rm osc}$ in fact evolve 
dynamically toward these stasis values, given the initial conditions we 
have specified for the $\phi_\ell$.  We shall perform our analysis of the cosmological 
dynamics by numerically solving the coupled system of equations of motion for
$H$ and $\phi_\ell$.  For the moment, we shall focus on the case in which 
$\Omega_{\rm tow}(t) = 1$ for all $t$ and defer discussion of the more general 
case until Sect.~\ref{sec:tracker}.

\begin{figure}[t!]
\includegraphics[width=0.48\textwidth]{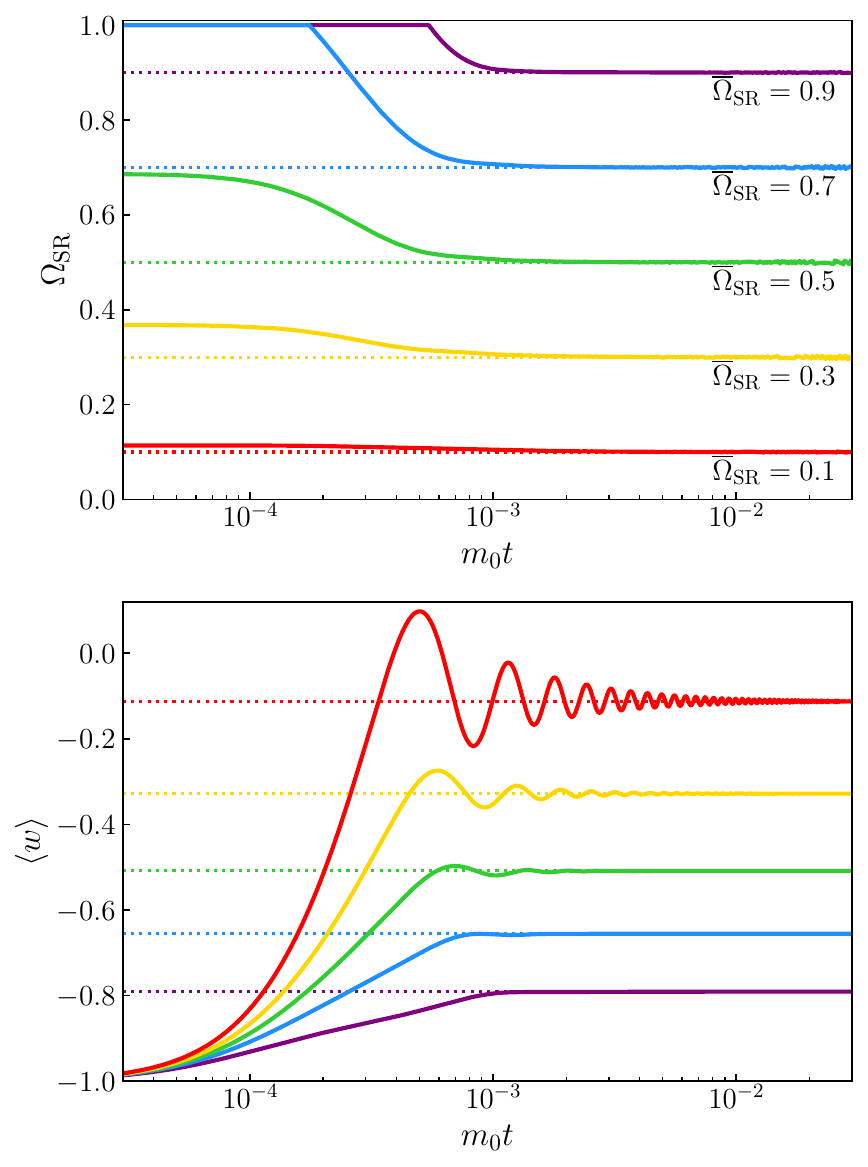}
  \caption{{\it Upper panel}\/: The abundances $\Omega_{\rm SR}(t)$ of the 
    slow-roll energy component (solid curves), plotted as functions of the 
    dimensionless time variable $m_0 t$ for 
    $H^{(0)}/m_{N-1} = \{0.23, 0.40, 0.55, 0.71, 0.97\}$.
    These values of $H^{(0)}/m_{N-1}$ respectively correspond to the stasis 
    abundances $\barOmega_{\rm SR} = \{0.1, 0.3, 0.5, 0.7, 0.9\}$.  For each curve, 
    the dotted horizontal line of the same color indicates the corresponding 
    value of $\barOmega_{\rm SR}$. 
    {\it Lower panel}\/: The effective equation-of-state parameters 
    $\langle w\rangle(t)$ for the scalar tower as a whole (solid curves), plotted 
    as functions of $m_0 t$ over the same range as in the upper panel.
    Each curve corresponds to the same value of $H^{(0)}/m_{N-1}$ as the curve of the 
    same color in the upper panel.  For each curve, the dotted horizontal line indicates the
    corresponding value of $\barw_{\rm SR}$.  All of the results shown in either panel 
    correspond to the parameter choices $\alpha = 1$ and $\delta = 1$.
\label{fig:examples}}
\end{figure}

In the upper panel of Fig.~\ref{fig:examples}, we plot the abundances 
$\Omega_{\rm SR}(t)$ of the slow-roll component (solid curves) as  functions of 
the dimensionless time variable $m_0 t$ for 
$H^{(0)}/m_{N-1} = \{0.23, 0.40, 0.55, 0.71, 0.97\}$.
The corresponding stasis abundances --- obtained from Eq.~(\ref{eq:Omegabar_condition}) 
with $\Omega_{\rm tow}^{(0)} = 1$ and with $\barkappa$ determined implicitly through 
Eq.~(\ref{eq:solve_kappa_1}) --- are respectively given by 
$\barOmega_{\rm SR} = \{0.1, 0.3, 0.5, 0.7, 0.9\}$.  For each 
$\Omega_{\rm SR}$ curve shown, the dotted horizontal line of 
the same color indicates the corresponding value of $\barOmega_{\rm SR}$. 
By contrast, in the lower panel of Fig.~\ref{fig:examples}, we plot the 
corresponding equation-of-state parameters $\langle w \rangle(t)$ for the 
tower as a whole (solid curves) as functions of $m_0 t$.  For each 
$\langle w\rangle$ curve shown, the dotted horizontal line of the 
same color indicates the corresponding value of $\barw_{\rm SR}$.  
All of the curves shown in Fig.~\ref{fig:examples} correspond to the 
parameter choices $\alpha = 1$ and $\delta = 1$.

We see from Fig.~\ref{fig:examples} that the universe evolves dynamically
toward stasis regardless of the initial value $H^{(0)}/m_{N-1}$.  However, 
consistent with our result in Eq.~(\ref{eq:Omegabar_condition}), we see that 
the particular stasis value $\barOmega_{\rm SR}$ towards which $\Omega_{\rm SR}$ 
evolves {\it does}\/ depend on this ratio.  We also see from this figure 
that the universe can remain in stasis, with  an effectively fixed abundance 
$\barOmega_{\rm SR}$, for a significant duration, even for a moderate value 
of $N$.  Indeed, the curves shown in Fig.~\ref{fig:examples} were calculated 
with $N=5000$, and even with this relatively small value the stases shown in
Fig.~\ref{fig:examples} have not yet reached their endpoints.  We shall 
discuss the relationship between $N$ and the resulting number of $e$-folds 
of stasis below.

We emphasize that while certain quantitative aspects of the $\Omega_{\rm SR}$ curves 
shown in Fig.~\ref{fig:examples}    reflect the particular values of $\alpha$ and $\delta$
we have chosen, the abundance curves obtained for other combinations of $\alpha$ 
and $\delta$ which likewise satisfy the stasis condition in Eq.~(\ref{eq:stasis_constraint})
are qualitatively similar.  Indeed, we find that the universe is generically 
attracted toward stasis in each case, despite the fact that $\barOmega_{\rm SR}$ 
depends on $H^{(0)}/m_{N-1}$.

More generally, we find that the universe is always attracted towards a stasis solution 
within this dynamical-scalar system {\it regardless}\/ of the initial conditions.   
The initial conditions affect the values of the abundances and equation-of-state 
parameters of our cosmological energy components during stasis, but the universe 
is always attracted towards a stasis configuration.

Since we are assuming that $\dot{\phi}_\ell^{(0)} = 0$ for all fields in the tower, 
we initially have $\expt{w}= -1$, regardless of the value of $H^{(0)}/m_{N-1}$.
Moreover, as the system evolves toward stasis, we also observe   that $\expt{w}$ 
approaches the constant value $\barw$ obtained from Eq.~(\ref{eq:w_sr}) with 
$\barkappa$ determined explicitly through Eq.~(\ref{eq:solve_kappa_1}).
In cases in which the initial value of $\Omega_{\rm osc}=1-\Omega_{\rm SR}$ is 
relatively large, we observe that the value of $\expt{w}$ oscillates around
$\barw_{\rm SR}$ before it settles into its stasis value.
This is due to the fact that the highly oscillatory $\phi_\ell$ have a more 
significant impact on the value of $\langle w \rangle$ when $\Omega_{\rm osc}$ is
large.  Indeed, this oscillatory behavior is less pronounced when $\Omega_{\rm SR}$ is 
relatively large and the contribution to $\expt{w}$ from $\Omega_{\rm osc}w_{\rm osc}$
is therefore less significant. 

As is the case with $\barOmega_{\rm SR}$ and $\barw$, we find that the duration of
the stasis epoch --- and therefore the number $\mathcal{N}_s$ of $e$-folds of expansion 
the universe undergoes during this epoch --- depends on the ratio $H^{(0)}/m_{N-1}$.
In the regime in which $3H^{(0)} > 2m_{N-1}$ and all of the $\phi_\ell$ are
effectively underdamped at $t=t^{(0)}$, we may obtain a rough estimate for $\mathcal{N}_s$ 
by approximating the duration of the stasis epoch to be the interval 
between the times $t_{N-1}$ and $t_0$ at which $\phi_{N-1}$ and $\phi_0$ undergo 
their critical-damping transitions, respectively.  Approximating 
$H \approx \barkappa/(3t)$ at each of these transition times and using the fact 
that the scale factor scales like $a\sim t^{\barkappa/3}$ during stasis, we find that
\begin{equation}
  \mathcal{N}_s~\approx~ \log\left[\frac{a(t_{0})}{a(t_{N-1})}\right]
    ~\approx~ \frac{\barkappa}{3}\log\left[\frac{t_{0}}{t_{N-1}}\right]\,.
  \label{eq:Ns_1}
\end{equation}
By contrast, in the opposite regime, in which $3H^{(0)} < 2m_{N-1}$, all 
$\phi_\ell$ with masses $m_\ell > 3H^{(0)}/2$ would begin oscillating immediately 
at $t =t^{(0)}$.  In this regime, then, $\mathcal{N}_s$ is given by a expression 
similar to that in Eq.~\eqref{eq:Ns_1}, but with $t_{N-1}\rightarrow t^{(0)}$.  
As a result, in either regime, we have
\beqn
  \mathcal{N}_s &\approx& 
  \frac{\barkappa}{3}
    \log\left(\frac{t_{0}}{\max\{t_{N-1},t^{(0)}\}}\right)\nn\\
  &\approx& \frac{\barkappa}{3}
    \Bigg[\delta\log N+\log\left(\frac{\Delta m}{m_0} \right)\nn\\
    & & ~~~~+ 
    \log\left(\frac{3H^{(0)}}{\max\{3H^{(0)},2m_{N-1}\}}\right)\Bigg]\,.~~~~~~
  \label{eq:Ns_2}
\eeqn

We see from Eq.~(\ref{eq:Ns_2}) that $\mathcal{N}_s$ increases 
logarithmically with the number of $\phi_\ell$ in the tower.  Moreover,
we also see that $\mathcal{N}_s$ increases monotonically with the ratio 
$H^{(0)}/m_{N-1}$ for a given choice of $m_0, \Delta m, \delta$, and $N$.  
This is due primarily to the fact that $\barkappa$ likewise increases 
monotonically with this ratio, but the growth of $\mathcal{N}_s$
with $H^{(0)}/m_{N-1}$ is also enhanced in the $H^{(0)}/m_{N-1} < 2/3$
regime due to the fact that all $\phi_\ell$ with $m_\ell > 3H^{(0)}/2$ 
begin oscillating immediately at $t = t^{(0)}$.

We note, however, that the expression in Eq.~(\ref{eq:Ns_2}) 
overestimates the value of $\mathcal{N}_s$ within the regime in which 
$\delta$ is small and the ratio $m_0/\Delta m$ is non-negligible.
Within this regime, the mass spectrum of the $\phi_\ell$ is significantly 
compressed and the value of $m_0$ therefore has a non-negligible impact on the 
$m_\ell$ across a significant portion of the tower.  As a result, although our 
fundamental scaling relation between $m_\ell$ and $\ell$ in Eq.~(\ref{eq:mass_spec}) 
continues to hold, this relation is no longer well approximated by the simpler 
power-law relation $m_\ell/\Delta m \approx \ell^\delta $ within this 
portion of the tower.  Thus,  the manner in which the $\phi_\ell^{(0)}$ scale with $\ell$ 
deviates significantly from the scaling relation in Eq.~(\ref{displacementscaling}).  
Thus, while the universe evolves toward and subsequently remains in stasis as long as
the total energy density of the tower remains dominated by $\phi_\ell$ with 
$m_\ell \gg m_0$, the stasis epoch effectively ends as soon as the lighter $\phi_\ell$ 
whose $\phi_\ell^{(0)}$ do not satisfy this scaling relation begin to dominate that 
total energy density.  That said, in the opposite regime, in which $m_0 \ll \Delta m$, 
the $\phi^{(0)}_\ell$ satisfy this scaling relation across essentially the entire 
tower, and the expression in Eq.~(\ref{eq:Ns_2}) still furnishes a reasonable 
estimate for $\mathcal{N}_s$.

\subsection{Alternative partitions}

Before moving forward, we comment on one additional property of our 
realization of stasis which bears mention.  The two cosmological energy 
components which coexist with constant abundances during stasis --- 
components which we have called ``slow-roll'' and ``oscillatory'' --- 
are each derived from collections of fields whose individual equation-of-state 
parameters evolve continuously from $w_\ell = -1$ to $w_\ell = 0$.
As indicated in Sect.~\ref{sec:preliminaries}, the criterion we have adopted 
in order to determine with which energy component a
given such field should be associated  at any particular time $t$ is
whether or not $3H(t) \geq 2m_\ell$.  If this criterion is satisfied for
a given $\phi_\ell$, we associate this field with the slow-roll component; if it
is not, we associate this field with the oscillatory component.
This is certainly a physically motivated choice, given that it 
associates all $\phi_\ell$ which are overdamped at time $t$ with the 
slow-roll component and all of the fields which are underdamped with the
oscillatory component.  

While the distinction between overdamped and underdamped fields is a mathematically 
important one, there is nevertheless no sharp distinction that occurs in the 
behavior of a given field as it crosses this boundary.  Even for a single scalar 
field evolving in a fixed external cosmology, as shown in Fig.~\ref{fig:phi_t}, 
the transition from the overdamped to underdamped regimes is a completely 
smooth one.  For this reason, it is natural to wonder whether our discovery of 
a stasis between the slow-roll and oscillatory components critically relies on 
this being taken as the definitional boundary between the two components, or 
whether an analogous stasis might exist even if this boundary were shifted in 
either direction.

As we shall now see, a stasis  emerges even if this boundary is shifted.  
More specifically, if we were to replace our standard ``slow-roll'' criterion 
$3H(t) \geq 2 m_\ell$ with a generalized criterion 
\beq
    3\, A\, H(t) ~\geq~ 2 \, m_\ell~
\label{Adef}
\eeq
where $A$ is an arbitrary positive constant, we would find 
that a stasis develops regardless of the value of $A$.  
Such a stasis would then take place between the abundances of 
the new ``slow-roll'' component (\ie, now defined as the component 
comprising fields which satisfy this modified criterion at time $t$) 
and the new ``oscillatory'' component (\ie, the component comprising 
fields which do not).

It is straightforward to understand why such a stasis continues to arise.
Since the upper limit of integration in Eq.~(\ref{eq:bessel_int_rho_V_prelim}) 
is given by $m_{\ell_c} = 3AH/2 = A\barkappa /(2t)$ for such a criterion, 
it follows that $I^{(\rho)}_{\rm SR}(\barkappa)$ is likewise independent of $t$ 
for any choice of $A$ and given by
\begin{equation}    
  I^{(\rho)}_{\rm SR}(\barkappa) ~=~ 
    \int_0^{A\barkappa/2}d\tt~\tt^{\alpha+1/\delta-\barkappa}
    \left[ J_{\frac{\barkappa+1}{2}}^2(\tt) 
    + J_{\frac{\barkappa-1}{2}}^2(\tt) \right] \,.
  \label{eq:bessel_int_gen_slice}
\end{equation}
The corresponding expressions for the quantity $I^{(P)}_{\rm SR}(\barkappa)$ 
Eq.~(\ref{eq:bessel_integral_P_V}) and for the quantities 
$I^{(\rho)}_{\rm osc}(\barkappa)$ and $I^{(P)}_{\rm osc}(\barkappa)$
are obtained by making the replacement $\barkappa/2 \rightarrow A\barkappa/2$ 
in the upper and lower limits of integration, respectively.  Since
since $I^{(\rho)}_{\rm SR}(\barkappa)$ and $I^{(\rho)}_{\rm osc}(\barkappa)$ 
are both constant during stasis for any choice of the partition parameter $A$, 
the corresponding abundances $\Omega_{\rm SR}$ and $\Omega_{\rm osc}$ are 
likewise constant during stasis whenever Eq.~(\ref{eq:stasis_constraint}) is 
satisfied.  Moreover, while the particular values $\barOmega_{\rm SR}$ and 
$\barOmega_{\rm osc}$ that these abundances take during stasis do depend 
non-trivially on $A$, we find that $\Omega_{\rm SR}$ and $\Omega_{\rm osc}$ 
evolve dynamically toward these stasis values for any choice of this 
partition parameter.

\begin{figure}
\includegraphics[width=0.49\textwidth]{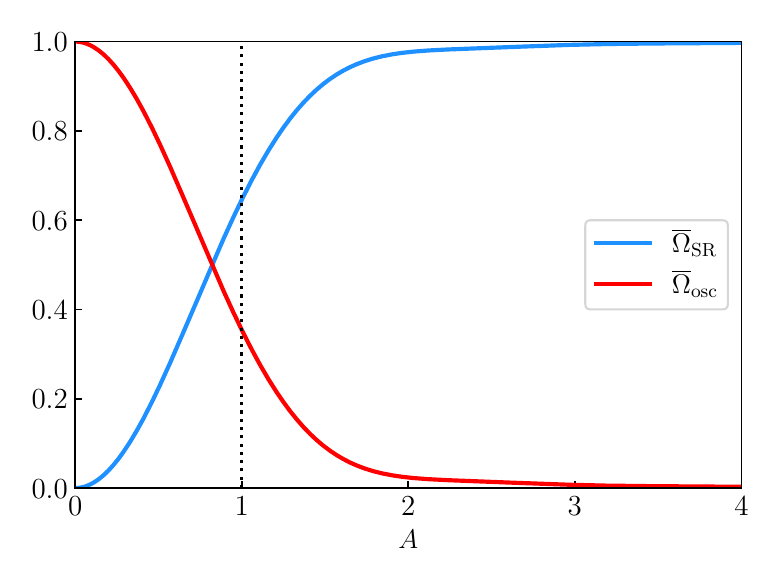}
  \caption{The stasis abundances $\barOmega_{\rm SR}$ and $\barOmega_{\rm osc}$ 
  that emerge for arbitrary choices of the partition parameter $A$ in 
  Eq.~(\ref{Adef}), plotted as functions of $A$ for the reference value 
  $H^{(0)}/m_{N-1}=2/3$.  The vertical line indicates the standard choice 
  $A=1$ that we have made throughout this paper.  We see that a stasis 
  emerges for all values of $A$ shown, with the resulting stasis abundances 
  increasingly favoring $\barOmega_\SR$ for larger $A$ and  $\barOmega_\osc$ 
  for smaller $A$.
\label{fig:Omega_kappa}}
\end{figure}

In Fig.~\ref{fig:Omega_kappa}, we plot the stasis abundances $\barOmega_{\rm SR}$ and 
$\barOmega_{\rm osc}$ as functions of $A$ for the parameter choice 
$H^{(0)}/m_{N-1}=2/3$.  For reference, we also include a dotted vertical 
line indicating our usual value $A=1$, which corresponds to choosing the 
partition location to coincide with the location  of the underdamping transition.   
As we see from this figure, the effect of increasing $A$ is to increase 
$\barOmega_\SR$ and to decrease $\barOmega_\osc$, with the opposite results 
arising for  decreasing $A$.  It is easy to understand this behavior.
Let us imagine increasing the value of $A$ during stasis.  This then effectively 
increases the critical value $\ell_c(t)$ of $\ell$ for which the criterion 
$3AH(t) \geq 2m_\ell$ is satisfied at any given time.
This in turn has the effect of shifting certain $\phi_\ell$ states from the 
set of states which contribute to $\barOmega_\osc$ to the set of states 
contributing to $\barOmega_\SR$.  This causes  the former abundance to decrease 
and the latter abundance to increase.  Moreover, this effect is never washed 
out at subsequent times because we are in stasis.  Thus the new re-partitioned 
abundances are fixed and do not evolve further.

Although we have seen that a stasis emerges over a wide range of values 
for $A$, there are intrinsic limits to how large or small $A$ may be taken.  
Indeed, these limits can be seen in Fig.~\ref{fig:Omega_kappa}: when $A$ is 
taken too large, $\barOmega_\osc$ falls to zero, while if $A$ is taken too small,
$\barOmega_\SR$ falls to zero.  Thus, in either extreme limit, we no longer 
obtain a meaningful stasis between two significant energy components.
We can also understand this behavior by thinking about the $\phi_\ell$ tower.
Given that we have posited a tower of $N$ components, it is possible for the 
value of $A$ to become so large or so small that we have either too few states 
in the oscillatory phase at the top of the tower at early times or too few 
states in the slow-roll phase at the bottom of the tower at late times.  
In either case, the prevalence of such significant ``edge'' effects can then 
prevent a stasis from developing at early times or surviving until late times.
These destructive effects arise because the existence of too few states in 
either scenario would invalidate some the approximations (such as the continuum 
approximation) that were made in Sect.~\ref{sec:ScalarTower}.~
This then seriously curtails (or potentially even completely eliminates) the 
length of time available for a corresponding stasis epoch.  However, as long 
as $A$ is not taken to these extremes, we see that we have a healthy stasis 
whose existence persists regardless of changes in $A$.

Thus far we have focused on the partitioning of our tower into only two energy components.   
However, we can even consider a more general partitioning of the tower into an arbitrary 
number $N_C$ of cosmological energy components.  These components may be labeled by
an index $i = 1, 2, \ldots, N_C$.  We define an abundance $\Omega_i$ and a 
partition parameter $A_i$ for each of these energy components such that $A_1=0$ 
and $A_{i+1} > A_i$.  For all $i < N_C$, we associate the abundances $\Omega_\ell$ 
of all of the $\phi_\ell$ with masses within the range 
$3A_{i+1}H(t) \geq 2m_\ell > 3A_i H(t)$ with $\Omega_i$.  We associate the 
abundances $\Omega_\ell$ of all the $\phi_\ell$ with masses 
$2m_\ell > 3 A_{N_C} H(t)$ with $\Omega_{N_C}$.  

For such a general partition, we find that when the condition in 
Eq.~(\ref{eq:stasis_constraint}) is satisfied, a stasis --- one in which {\it all}\/ 
of the $\Omega_i$ take effectively constant values --- likewise emerges in the continuum 
limit.  These stasis abundances $\barOmega_i$ are given by functions of the form in 
Eq.~(\ref{eq:Omega_sr_1}) with $I^{(\rho)}_{\rm SR}(\barkappa)$ replaced by a function 
of the form
\begin{equation}
  I^{(\rho)}_{i}(\barkappa) ~=~ 
    \int^{A_{i+1}\barkappa/2}_{A_i\barkappa/2}d\tt~\tt^{\alpha+1/\delta-\barkappa}
    \left[ J_{\frac{\barkappa+1}{2}}^2(\tt) 
    + J_{\frac{\barkappa-1}{2}}^2(\tt) \right]
\end{equation}
for $i < N_C$, and by a similar expression with the replacement 
$A_{i+1}\barkappa/2\rightarrow \infty$ in upper limit of integration for $i = N_C$.
We also find numerically that the $\Omega_i$ are dynamically attracted toward their 
corresponding $\barOmega_i$ values for arbitrary such partitionings of the tower into 
energy components.

We see, then, that the emergence of stasis from a tower of dynamical scalars does not
depend on the manner in which we partition the tower into energy components based 
on the relationships between $H(t)$ and the individual $m_\ell$.  That said,
the two-component partition which we have been employing thus far in this paper,
in which one component comprises the scalars which are overdamped at any given
time and the other comprises the scalars which are underdamped, is a physically
meaningful one, and we shall continue to adopt this partition in what follows.

\FloatBarrier


\section{Stasis in the presence of a background energy component\label{sec:tracker}}


In this section we investigate what happens if we repeat our previous analysis, only 
now in the presence of an additional energy component which we may regard as a 
``background spectator'' --- \ie, a fluid which is completely inert, neither 
receiving energy from our $\phi_\ell$ fields nor donating energy to them. 
We shall conduct our analysis in two stages.  First, we will consider the case 
in which this background is time-independent, with a fixed equation-of-state 
parameter $w_\BG$.   We shall then consider how our results are modified if our 
background has a time-dependent equation-of-state parameter $w_\BG(t)$.

\subsection{Time-independent background}

We begin our analysis by considering the case in which our background fluid has 
a fixed equation-of-state parameter  $w_{\rm BG}$.  We shall make no other 
assumptions regarding the nature of this background and we shall allow its 
initial abundance $\Omega^{(0)}_{\rm BG}$ to be completely arbitrary.  Thus, 
even though we shall refer to this energy component as a ``background'', we 
shall {\it not}\/ assume that it dominates the cosmology of our system.

In the following analysis, we shall let $\overline{w}$ represent the 
equation-of-state parameter of our dynamical-scalar system {\it during the 
stasis that would have resulted if there had been no extra background 
component}\/.  Indeed, $\overline{w}$ will continue to be given by 
Eq.~\eqref{first_time_again} where $\barOmega_\SR$ and $\barOmega_\osc$ 
are likewise the values of $\Omega_\SR$ and $\Omega_\osc$
that would have emerged in such a background-free stasis.  As long as our slow-roll and 
oscillatory components have reached stasis, all of the quantities in 
Eq.~\eqref{first_time_again} are time-independent.  We shall also continue to let 
$\langle w\rangle$ denote the time-dependent equation-of-state parameter for our $\phi_\ell$ 
tower alone, as in Eq.~(\ref{eq:w_eff_all}).  By contrast, we shall let $w_{\rm univ}$ 
continue to denote the equation-of-state parameter for the entire universe, bearing in 
mind that this now includes not only the contribution from the $\phi_\ell$ tower but 
also the contribution from the background:
\beq
   w_{\rm univ} ~\equiv~
   \sum_i w_i \Omega_i ~=~
   \langle w\rangle \,\Omega_{\rm tow} + 
   w_\BG \,\Omega_\BG~.
\label{eq:kappabar_bg}
\eeq
Indeed, with this definition Eq.~(\ref{eq:kappa_univ}) continues to apply.

As we have already remarked at the end of Sect.~\ref{subsec:stasis_condition}, 
we do not expect the {\it existence}\/ of a stasis solution to be disturbed 
by the introduction of a background. However, what interests us here are the 
answers to two questions:
\begin{itemize}
  \item How is the stasis solution affected by the presence of the spectator 
    background?
  \item How is the {\it dynamics}\/ of our system affected by the presence of 
    the spectator background?  Does the (possibly new) stasis solution 
    continue to serve as an attractor?
\end{itemize}
In this section, we shall provide answers to these questions.

\begin{figure*}
\includegraphics[width=0.99\textwidth]{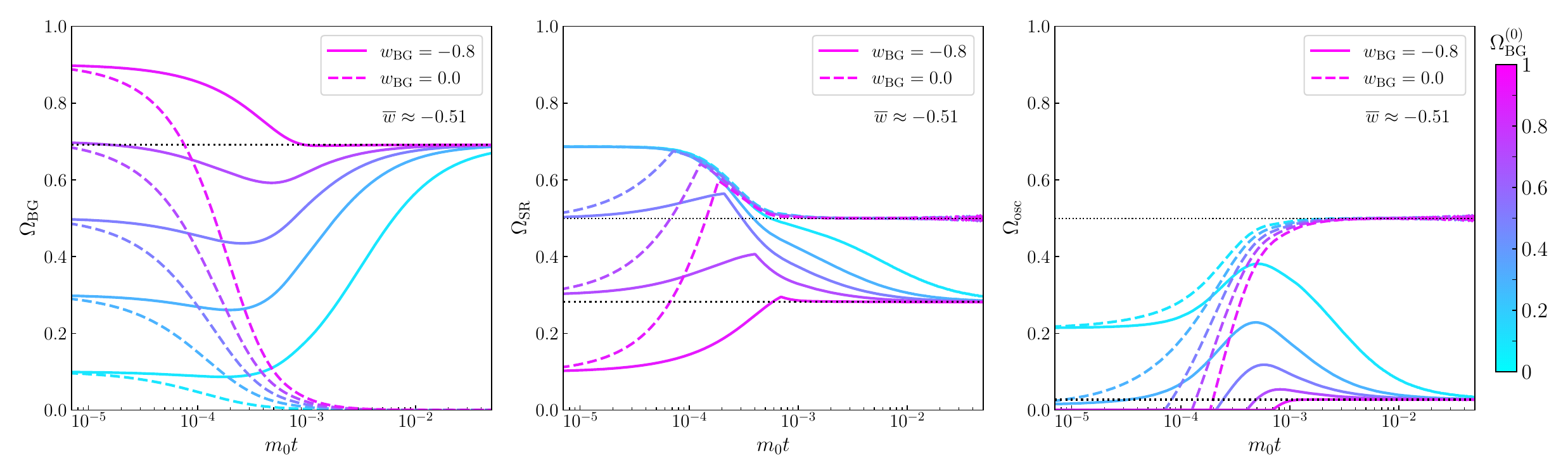}
\caption{The time-evolution of the abundances $\Omega_\BG$ (left panel), $\Omega_\SR$ 
(middle panel), and $\Omega_\osc$ (right panel), plotted as functions of $m_0 t$ 
for two different choices of the equation-of-state parameter $w_\BG$ for
the background fluid (solid versus dashed lines).  The differently colored lines 
represent different choices of the initial abundance $\Omega_\BG^{(0)}$ for this
background fluid.  The parameters chosen for these plots correspond to a system 
which would have had $\barw= -0.51$ in the absence of the background fluid.  
In all cases, we see that the system evolves towards a stasis solution in which 
$\Omega_\SR$ and $\Omega_\osc$ have constant, non-zero values --- values which 
are independent of $\Omega_\BG^{(0)}$ but nevertheless depend on
the value of 
$w_\BG$.  Indeed, for $w_\BG=0$ (dashed lines), we observe that $\Omega_\BG\to 0$
as time increases for all $\Omega_\BG^{(0)}$.  By contrast, for $w_\BG= -0.8$ 
(solid lines), we find that $\Omega_\BG$ always asymptotes to a 
fixed non-zero value.}  
\label{fig:abundances_BG}
\end{figure*}

\begin{figure}
\includegraphics[width=0.45\textwidth]{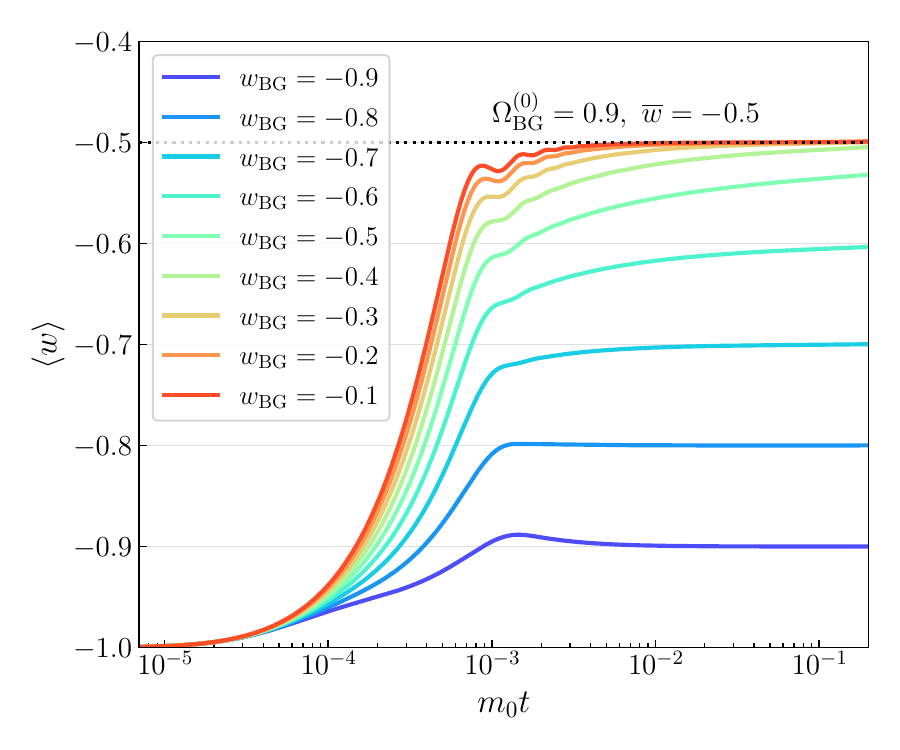}
\caption{The effective equation-of-state parameter $\langle w\rangle$ for a tower 
with $\overline{w}= -0.5$, evaluated in the presence of a background fluid with 
equation-of-state parameter $w_\BG$ and plotted as a function of time for different 
fixed values of $w_\BG$.  As we sweep through increasing values of $w_\BG$, we find 
that $\langle w\rangle$ always reaches a stasis value.  For $w_\BG > \overline{w}$, 
we find that this stasis value saturates at $\overline{w}$.  By contrast, for 
$w_\BG < \overline{w}$, we find that this stasis value is given by $w_\BG$.
Thus, for $w_\BG < \overline{w}$, we see that the stasis equation-of-state 
parameter $\langle w\rangle$ for our system 
{\it tracks}\/ that of the 
background.}
\label{fig:w_evolution_BG}
\end{figure}

We begin by investigating the effect that varying both the initial abundance
$\Omega_{\rm BG}^{(0)}$ and equation-of-state parameter $w_{\rm BG}$ of the 
background has on the manner in which the abundances $\Omega_\SR$ and 
$\Omega_\osc$ evolve with time.  In Fig.~\ref{fig:abundances_BG}, we 
plot $\Omega_{\rm BG}$ (left panel), $\Omega_{\rm SR}$ (middle panel), 
and $\Omega_{\rm osc}$ (right panel) as functions of $m_0 t$ for several different 
combinations of $\Omega_{\rm BG}^{(0)}$ and $w_{\rm BG}$.

For all combinations of $\Omega_{\rm BG}^{(0)}$ and $w_{\rm BG}$, we 
observe that the universe evolves towards a stasis in which $\Omega_\SR$ and 
$\Omega_\osc$ have constant, non-zero values.  In this three-component system,
this of course implies that $\Omega_\BG$ evolves toward a constant value as well.
Moreover, we observe that the value of $\Omega_{\rm BG}^{(0)}$ has no effect on
the constant values toward which $\Omega_{\rm SR}$ and $\Omega_{\rm osc}$ 
ultimately evolve.  Indeed, the stasis that emerges for a given choice of 
$w_{\rm BG}$ is also completely independent of $\Omega_\BG^{(0)}$.
This is already an interesting result --- one which confirms our expectation 
that the presence of a background component should affect neither the 
{\it existence}\/ of a stasis solution within our dynamical system, nor the 
fact that this solution is an attractor within that system. 

That said, we also see from Fig.~\ref{fig:abundances_BG} that the stasis which 
emerges in the presence of a background component depends non-trivially on the 
value of $w_\BG$.  The results shown for the larger value of $w_\BG$ (dashed lines) 
indicate that $\Omega_\BG\to 0$ for all choices of $\Omega_{\rm BG}^{(0)}$.  
Furthermore, the stasis abundances which ultimately emerge for the slow-roll and 
oscillatory components after the background abundance dies away are precisely the 
same stasis abundances that we would have obtained for a slow-roll/oscillatory-component 
stasis with background absent altogether.  By contrast, the results for the smaller 
value of $w_\BG$ (solid lines) indicate that $\Omega_\BG$ asymptotes toward a finite, 
non-zero value.  Thus $\barOmega'_\SR + \barOmega'_\osc < 1$ for the stasis that emerges, 
where $\barOmega_{\rm SR}'$ and $\barOmega'_{\rm osc}$ denote the modified stasis 
abundances for the slow-roll and oscillatory energy components which emerge in the 
presence of the background. 

In order to further elucidate the manner in which $\barOmega_{\rm SR}'$ and 
$\barOmega'_{\rm osc}$ depend on $w_\BG$, in Fig.~\ref{fig:w_evolution_BG} 
we plot $\langle w\rangle$ as a function of $m_0t$ for a variety of 
different choices of $w_\BG$.  All curves shown correspond to the parameter 
choices $\overline{w} = -0.5$ and $\Omega_{\rm BG}^{(0)} = 0.9$.
For $w_\BG < \overline{w}$, we find that the stasis value of $\langle w\rangle$ 
is given by $w_\BG$.  By contrast, for $w_\BG > \barw$, we 
find that this stasis value saturates at $\overline{w}$. 

For $w_\BG > \overline{w}$, this latter result is easy to understand.   
As our system evolves, $\overline{w}$ is less than $w_\BG$.  Thus our background 
redshifts away, \ie,
\beq 
  \Omega_\BG ~\to~ 0~,
\eeq
purely as a consequence of cosmological expansion.  Indeed we have already seen
this behavior within the dashed curves of the left panel of 
Fig.~\ref{fig:abundances_BG}.  Thus our system is ultimately attracted to  
the same stasis configuration as we would have had if the background had never 
been present, with $\Omega_\SR\to \barOmega_\SR$ and $\Omega_\osc\to \barOmega_\osc$.
It is for this reason that $\langle w\rangle\to \overline{w}$.
The stasis values for the abundances that emerge in this case are nothing but 
the values that are predicted by replacing $H^{(0)}$ with 
$H^{(0)}\sqrt{\Omega_{\rm tow}^{(0)}}$ in Eq.~\eqref{eq:solve_kappa_1}.
In other words, we reproduce our original stasis that emerged in the absence 
of a background but with the same total initial energy density of the tower.
This makes sense,  since there is no background energy component remaining in 
the system.  In such cases, the earlier period during which the background energy 
component still exists can then be viewed as a ``pre-history'' to the overall story.

By contrast, the manner in which the abundances behave for $w_\BG < \overline{w}$ 
is completely different.  Within this regime, our background does not redshift away, 
and indeed $\Omega_\BG$ asymptotes to a non-zero stasis value.   This means that 
$\Omega_\SR$ and $\Omega_\osc$ can no longer asymptote to the same stasis values 
that they would have had if no background had been present.  In other words, in 
this case {\it the presence of the background necessarily deforms the stasis 
away from what it would have been if the background had not been present.}\/  
Remarkably, however, the new stasis that is realized is one wherein 
\beq 
  \overline{w}' ~=~ w_\BG~,
\label{tracking}
\eeq
where $\barw'$ denotes the modified value which $\langle w\rangle$ takes during
stasis in the presence of the background.  In other words, the new stasis that is 
realized in this case has an equation-of-state parameter which {\it tracks}\/ that 
of the background!  This tracking behavior occurs 
regardless of the initial abundance $\Omega_\BG^{(0)}$ of the background --- indeed, 
the background need not even be dominant.  Moreover, this behavior occurs regardless 
of the value that $w_\BG$ takes, so long as $w_\BG < \overline{w}$.

This, then, is our first example of a ``tracking'' stasis.  As long as a background 
is present, and as long as $w_\BG < \overline{w}$ (so that this background survives 
into the stasis epoch without redshifting away), the modified equation-of-state 
parameter $\barw'$ for the tower will always match that of the background. 
If the initial background abundance $\Omega_\BG^{(0)}$ is large, this deformation 
of our stasis solution to match the background occurs relatively quickly. 
For smaller $\Omega_\BG^{(0)}$, by contrast, the deformation occurs more slowly.  
However, so long as $\Omega_\BG^{(0)}\not=0$, the abundances of the three 
cosmological energy components in our system will automatically reconfigure 
themselves such that $\barw'$ matches $w_\BG$.

\begin{figure*}
\includegraphics[width=0.75\textwidth]
{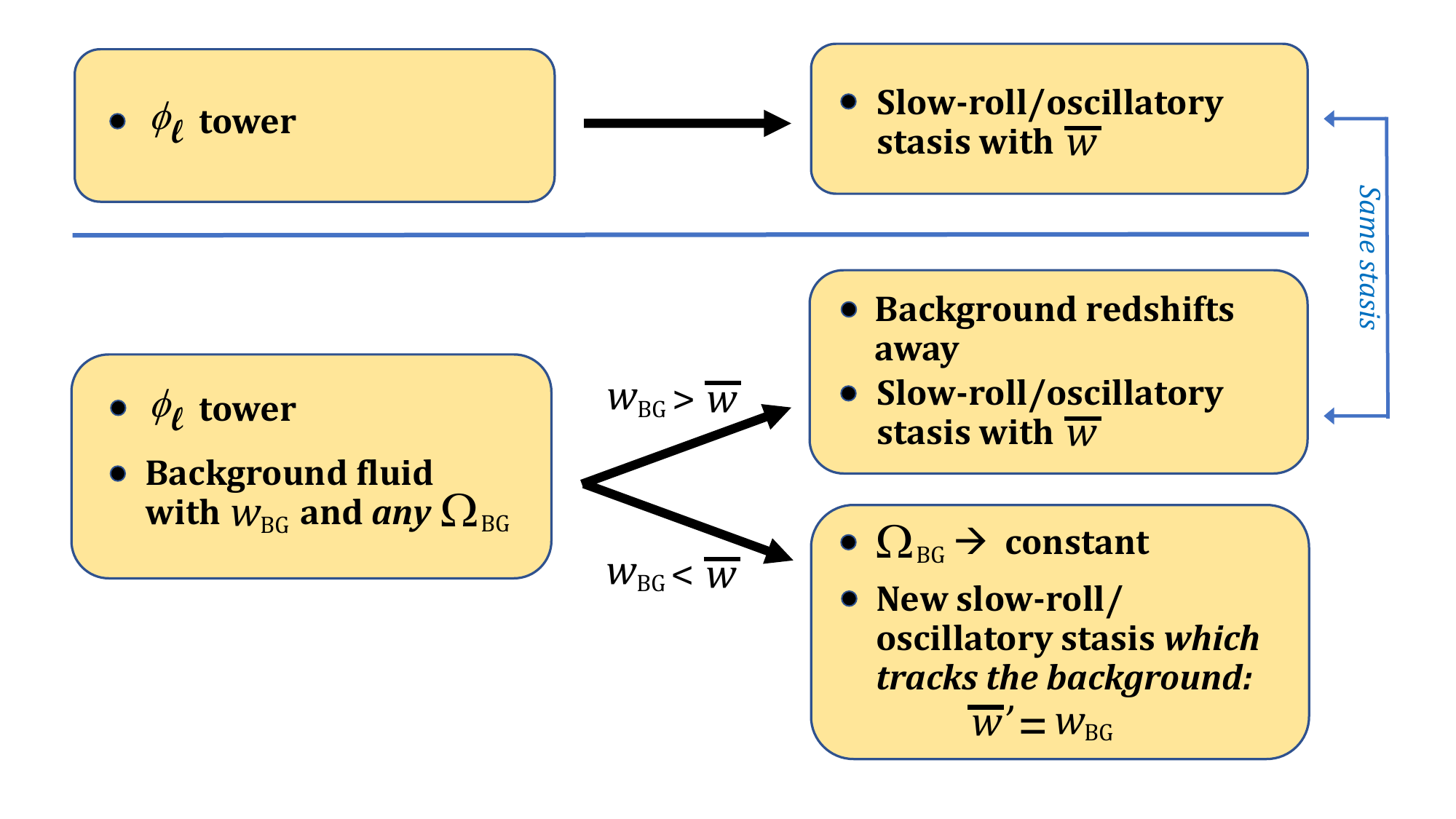}
    \caption{The cosmological evolution of a tower of scalar fields $\phi_\ell$ 
    in the presence of a background with equation-of-state parameter $w_{\rm BG}$.
    Without the background (top row of figure), we have seen in 
    Sect.~\ref{sec:ScalarTower} that our system naturally evolves into a 
    slow-roll/oscillatory-component stasis with an equation-of-state parameter 
    $\overline{w}$.   Given this, the consequences of introducing the 
    background (lower row of figure) depend on how $w_{\rm BG}$ relates to 
    $\overline{w}$.  If $w_{\rm BG}> \overline{w}$, the background simply 
    redshifts away while the tower continues to produce the same 
    slow-roll/oscillatory-component stasis as before.   However, if 
    $w_{\rm BG} < \overline{w}$, we find that $\Omega_{\rm BG}$ evolves towards 
    a non-zero constant while the tower now evolves into a {\it different}\/ 
    stasis configuration wherein the modified equation-of-state parameter 
    $\overline{w}'$ is equal to $w_{\rm BG}$.  In this way, then, the 
    equation-of-state parameter for the tower during stasis {\it tracks}\/ the 
    background. 
    \label{fig:tracker}}
\end{figure*}

Given these observations, the cosmological evolution of a tower of scalar fields 
$\phi_\ell$ in the presence of a background fluid  with equation-of-state parameter 
$w_{\rm BG}$ can be summarized as in Fig.~\ref{fig:tracker}.  The top part of this 
figure indicates that our system {\it without}\/ any background produces a stasis with 
a certain equation-of-state parameter $\overline{w}$.   The lower part of the figure 
then illustrates that when the background is present, we obtain a stasis whose 
equation-of-state parameter is generally given by
\beq
\langle w \rangle ~\to ~
  \min\left\{\overline{w}, \, w_{\rm BG} \right\}~.
\eeq 
Thus, for $w_\BG< \overline{w}$, we obtain Eq.~(\ref{tracking}).

This result makes perfect sense.  In Ref.~\cite{Dienes:2021woi} we demonstrated 
on general grounds that a pairwise stasis cannot exist in the presence of a 
spectator unless the two parts of the system --- \ie, the tower of dynamical scalars 
and the background spectator --- have the same effective equation-of-state parameter.  
It is thus the result in Eq.~(\ref{tracking}) that enables a stasis between 
the slow-roll and oscillatory components within the tower to arise even in 
the presence of the background spectator.  Indeed, we see that the only way in 
which we can evade having $\overline{w}' = w_\BG$ is to have $w_\BG < \overline{w}$.   
In that regime, the background spectator redshifts away, simply leaving us 
with $\overline{w}'=\overline{w}$.

At this stage, several comments are in order.  First, even though $\Omega_\BG$, 
$\Omega_\SR$, and $\Omega_\osc$ all asymptote to fixed, non-zero values for 
$w_\BG<\overline{w}$, it is important to note that this is not a triple stasis 
of the sort discussed in Ref.~\cite{Dienes:2023ziv}.  Indeed, in this scenario,
there is no transfer of energy between the background energy component and either
$\rho_{\rm SR}$ and/or $\rho_{\rm osc}$.  Rather, the behavior which our system
exhibits for $w_\BG < \overline{w}$ exemplifies a possibility discussed in 
Ref.~\cite{Dienes:2021woi}, wherein which a pairwise stasis between 
two cosmological components takes place in the presence of a background 
component.   

Second, we remark that it is only this slow-roll/oscillatory-component 
stasis achieved through dynamical scalars which has the ability to track 
a background.  This does not happen for any realization of stasis previously 
identified in the literature, including the matter/radiation stasis outlined 
in Refs.~\cite{Dienes:2021woi,Dienes:2022zgd} or {\it any}\/ of the pairwise 
stases --- or even the triple stasis --- discussed in Ref.~\cite{Dienes:2023ziv}.
The underlying reason for this is actually quite simple.  In all of these
other realizations of stasis, the underlying constraint equations relate 
$\alpha+1/\delta$ directly to the value of the equation-of-state parameter 
$\barw$ during stasis.   For example, in the case of matter/radiation stasis 
achieved via towers of decaying matter fields~\cite{Dienes:2021woi}, our 
underlying constraint equation took the form
\beq
  \frac{1}{\gamma}\left( \alpha+1/\delta\right) ~=~ 
    2- \barkappa ~=~ \frac{2 \overline{w}}{1+\overline{w}}~,
\eeq
where $\gamma$ is a parameter governing the scaling of decay widths $\Gamma_\ell$ 
across the tower and where $\barkappa= 2/(1+\overline{w})$.  Thus, once one 
specifies the fixed underlying parameters $(\alpha,\gamma,\delta)$ of our model,
the resulting stasis value $\overline{w}$ is fixed and cannot be altered.   This 
implies that if we introduce a spectator with equation-of-state parameter 
$w_{\rm BG}$ into such systems, and if $w_{\rm BG}$ differs from the value 
of $\barw$ predicted from the constraint equations, there is no mechanism via 
which the value of $\barw$ can be deformed such that it matches $w_{\rm BG}$.   
In other words, within the stasis systems discussed in 
Refs.~\cite{Dienes:2021woi,Dienes:2023ziv}, the components involved in the 
stasis do not have the freedom needed in order to track the spectator.

By contrast, the slow-roll/oscillatory-component stasis that we are discussing in this paper 
rests upon the much simpler constraint equation in Eq.~(\ref{eq:stasis_constraint}).   
Indeed, this constraint equation does not involve $\barw$ at all, which means that fixing 
the underlying model parameters $(\alpha,\gamma,\delta)$ does not fix a unique value for $\barw$.
This in turn means that the properties of the stasis within our dynamical-scalar scenario 
can be adjusted --- even to the extent of changing the value of the equation-of-state 
parameter $\overline{w}$ --- while still satisfying our underlying stasis condition.  
Thus, we see that {\it it is only the slow-roll/oscillatory-component stasis achieved 
through dynamical scalars which has the freedom needed to ``track'' a spectator field}\/.  
This feature thereby distinguishes this stasis from all of the stases that have previously 
been discussed in the literature.  This tracking phenomenon may have important
implications when this stasis is embedded in specific cosmological contexts~\cite{toappear}. 

Thus far, our discussion in this section has been primarily qualitative, based on the 
numerical results in Figs.~\ref{fig:abundances_BG} and \ref{fig:w_evolution_BG}.
However, it is not difficult to understand all of these features at an algebraic level.
For example, given that our universe contains two energy components (the tower of 
$\phi_\ell$ states and the background), it follows from Eq.~(\ref{eq:kappabar_bg}) 
that 
\begin{equation}
   w_{\rm univ} ~\equiv~
    \left( \langle w\rangle - w_\BG\right)\Omega_{\rm tow}  + w_\BG~.
  \label{eq:kappabar_bg_2}
\end{equation}
We likewise know that $\rho_{\rm tow}\sim a^{-3(1+ \overline{w}')}$
and $\rho_{\rm BG}\sim a^{-3(1+ w_{\rm BG})}$ during stasis,
whereupon we see that 
\beq
  \barOmega'_{\rm tow} = 0 ~{\rm or}~1~~~
    {\rm unless}~~ \overline{w}' = w_{\rm BG}~.
  \label{constraint2}
\eeq
Bearing in mind that within Eq.~(\ref{eq:kappabar_bg_2}) only 
$\langle w\rangle$ and $\Omega_{\rm tow}$ are time-dependent, we can then 
seek to determine the general conditions under which $w_{\rm univ}$ is 
time-independent.  It turns out that there are only three ways in which we may 
obtain a constant equation-of-state parameter $w_{\rm univ}$ for the universe as 
a whole while maintaining consistency with Eq.~(\ref{constraint2}):
\begin{itemize}
  \item the ``trivial'' solution without the tower, with $\Omega_{\rm tow}=0$, 
    whereupon we trivially have $w_{\rm univ} = w_{\rm BG}$;  
  \item  the original stasis without the background, with
    $\Omega_{\rm tow}=1$ and $\langle w\rangle=\overline{w}$, 
    leading to $w_{\rm univ}= \overline{w}$;  and
  \item the solution in which the tower has reached stasis with 
    $\langle w \rangle = w_\BG$, whereupon $w_{\rm univ} = \overline{w}' = w_\BG$.
\end{itemize}
Indeed, of these three solutions, the first is trivial while the second corresponds 
to the situation described in Fig.~\ref{fig:tracker} with $w_\BG > \overline{w}$ 
and the third corresponds to the ``tracking'' situation described in 
Fig.~\ref{fig:tracker} with $w_\BG < \overline{w}$.

We can also obtain explicit expressions for the abundances during such a tracking stasis.
Indeed, for $w_{\rm BG} < \barw$, Eq.~(\ref{eq:kappabar_bg_2}) reduces during stasis to
\begin{equation}
   \barw_{\rm univ}' ~\equiv~
    \left( \barw' - w_\BG\right)\barOmega_{\rm tow}' + w_\BG~.
  \label{eq:kappabar_bg_2_in_stasis}
\end{equation}
We can obtain an independent relation between $w_{\rm BG}$ and $\barOmega_{\rm tow}'$ 
from our general expressions in Eq.~\eqref{eq:Omegabar_condition} 
and~\eqref{eq:Omegabar_osc_codecil} for the stasis abundances of the 
slow-roll and oscillatory components of the tower, respectively --- expressions
which hold regardless of whether the background is present.  However,
since this relation holds in general, irrespective of the relationship between 
$w_{\rm BG}$ and $\barw$, we define $\barOmega_{\rm tow}^\ast$ to represent the total 
stasis abundance of the tower states in the presence of a background with a 
completely arbitrary value of $w_{\rm BG}$ --- a total abundance which may be 
given by either $\barOmega_{\rm tow}$ or $\barOmega_{\rm tow}'$, depending on 
circumstances.  Similarly, we define $\starbarkappa$ to represent a completely 
arbitrary value of $\kappa$ during stasis, which may likewise be given by either 
$\barkappa$ or $\barkappa'$.  
\beq
  \barOmega_{\rm tow}^\ast~=~ 
    \frac{18\left[I_{\rm SR}^{(\rho)}(\starbarkappa)+
      I_{\rm osc}^{(\rho)}(\starbarkappa)\right]}
    {\starbarkappa^2\mathcal{J}(\starbarkappa)}
    \left(\frac{H^{(0)}}{m_{N-1}}\right)^2\Omega_{\rm tow}^{(0)}~.
\label{eq:Omegabartower_bg}
\eeq
The abundance of the tower within the $w_{\rm BG} < \barw$ and $w_{\rm BG} > \barw$ 
regimes are obtained by taking $\starbarkappa = \barkappa'\equiv 2/(1+w_{\rm BG})$ 
and $\starbarkappa = \barkappa$ in the above equation, respectively.

Within the $w_{\rm BG} < \barw $ regime, we may solve Eqs.~\eqref{eq:kappabar_bg_2_in_stasis} 
and~\eqref{eq:Omegabartower_bg} together numerically in order to obtain
values for $\starbarkappa = \barkappa'$.  There exist two solutions to this system of equations: 
one which yields $\{\barOmega_{\rm tow}=1,\, \barOmega_{\rm BG}= 0,\, \expt{w}=\barw\}$ and one
which yields $\{\barOmega'_{\rm tow}<1,\, \barOmega_{\rm BG}>0,\, \expt{w}=w_{\rm BG} \}$.
By contrast, within the $w_{\rm BG} > \barw$ regime, the equation-of-state parameter 
for the universe during stasis is given by $\barw_{\rm univ} = \barw$, and the only 
physically consistent solution for $\starbarkappa = \barkappa$ is the one with 
$\{\barOmega_{\rm tow}=1, \,\barOmega_{\rm BG}= 0,\, \expt{w}=\barw\}$.  The reason that
a second solution for $\starbarkappa$ does not arise in this case ultimately owes to 
the fact that Eq.~\eqref{eq:Omegabartower_bg} is a decreasing function of 
$\starbarkappa$.  As a result, the would-be solution for $\starbarkappa$ that would 
otherwise have be obtained by taking $\barw \rightarrow w_{\rm BG}$ yields a value 
of $\barOmega_{\rm tow}$ which exceeds unity and must therefore be discarded.

We can use these results in order to obtain the individual abundances of our
slow-roll and oscillatory components.  In order to do this, we first note that the 
fraction of the abundance associated  with the slow-roll component within the 
tower during stasis is simply
\beq
 \frac{\barOmega_{\rm SR}^\ast}
   {\barOmega_{\rm osc}^\ast+\barOmega_{\rm SR}^\ast} ~=~ 
   \frac{I_{\rm SR}^{(\rho)}(\starbarkappa)}
   {I_{\rm osc}^{(\rho)}(\starbarkappa)+I_{\rm SR}^{(\rho)}(\starbarkappa)}~,
\eeq
which is only a function of $\starbarkappa$.
After obtaining the solution for $\barkappa'$ and $\barOmega_{\rm tow}'$, 
it is easy to see that $\barOmega_{\rm SR}'$ takes the same form as 
Eq.~\eqref{eq:Omegabar_condition}.  However, since the value of $\starbarkappa$ 
has changed from $\barkappa$ to $\barkappa'$ in the presence of the background,
we see that the abundance associated with the slow-roll component is modified to
\beq
  \barOmega_{\rm SR}' ~=~ \left[ 
    \frac{\barkappa^2\mathcal{J}(\barkappa)}{\barkappa'^2\mathcal{J}(\barkappa')}
    \frac{I_{\rm SR}^{(\rho)}(\barkappa')}{I_{\rm SR}^{(\rho)}(\barkappa)}\right]
    \barOmega_{\rm SR}~.
\eeq
The corresponding expression for $\barOmega_{\rm osc}'$ is likewise modified to
\beq
  \barOmega_{\rm osc}' ~=~ \left[ 
    \frac{\barkappa^2\mathcal{J}(\barkappa)}{\barkappa'^2\mathcal{J}(\barkappa')}
    \frac{I_{\rm osc}^{(\rho)}(\barkappa')}{I_{\rm osc}^{(\rho)}(\barkappa)}\right]
    \barOmega_{\rm osc}~.
\eeq
Indeed, these expressions accord with the modifications of the stasis 
abundances apparent in the numerical results shown in 
Fig.~\ref{fig:abundances_BG}.

\subsection{Time-dependent backgrounds and tracking solutions}

\begin{figure}
\includegraphics[width=0.45\textwidth]{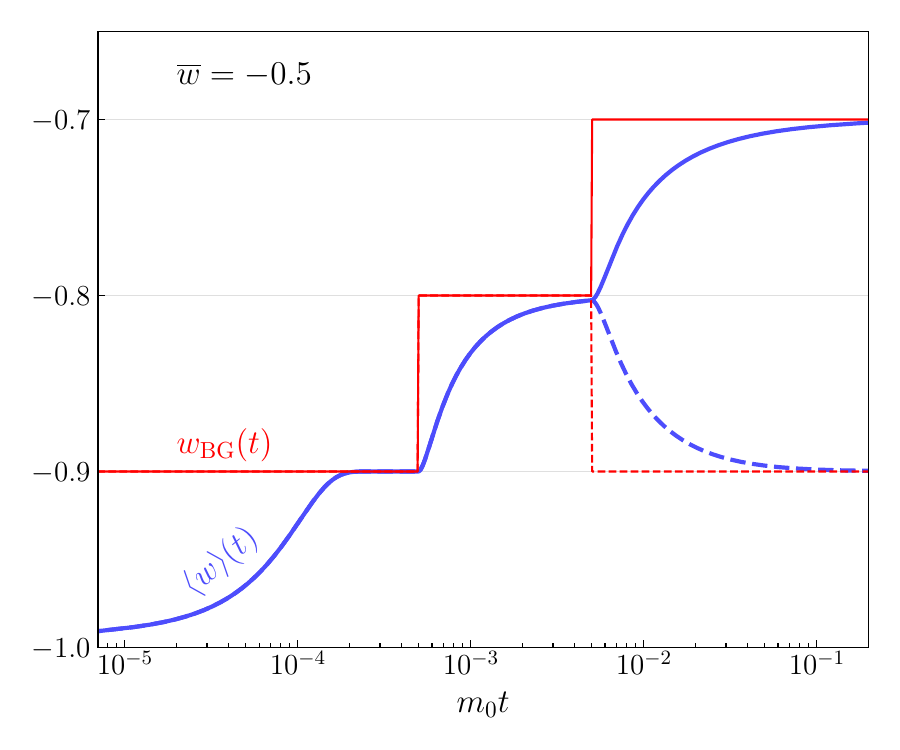}
\caption{Tracking behavior for the stasis state resulting from a tower of 
scalar fields in the presence of a {\it variable}\/ background.  Even when a 
stasis is achieved for the tower of scalar fields, any subsequent 
change in $w_{\rm BG}$ destabilizes the existing stasis and produces a new 
stasis for which $\langle w\rangle$ continues to match $w_{\rm BG}$.  This 
behavior persists as long as the new value of $w_{\rm BG}$ continues to be 
smaller than $\overline{w}$.  In this figure, two sets of changes for 
$w_{\rm BG}$ are shown: in one case (shown in solid red), the background 
has a value of $w_{\rm BG}$ which starts at $-0.9$ and subsequently jumps 
to $-0.8$ and then $-0.7$, while  in the other case (dashed red), the 
background behaves identically except that the final jump is back down 
to $w_{\rm BG} = -0.9$ rather than to $-0.7$.  
In both cases, the values of $\langle w\rangle$ for our scalar-field 
tower (shown in blue) attempt to track the behavior of the background, 
achieving short-lived stases with $\langle w\rangle = w_{\rm BG}$ within 
each interval before becoming destabilized again. The blue curves in this 
figure are calculated assuming an initial background abundance 
$\Omega_{\rm BG}^{(0)}=0.98$, but the same qualitative results would 
emerge for any value of $\Omega_{\rm BG}^{(0)}$ and any sequence of 
$w_{\rm BG}$-values which are all smaller than $\overline{w}$.
\label{fig:w_track}}
\end{figure}

Let us now consider what happens when the equation-of-state parameter 
$w_\BG$ is time-dependent.   We have already seen that when $w_\BG$ is a 
constant, and when this constant is smaller than $\overline{w}$,
the dynamics of our $\phi_\ell$ tower adjusts in order to realize a stasis 
with $\overline{w}'=w_\BG$.  It is therefore important to understand how 
our system responds when $w_\BG$ itself is changing.

In general, the time variation of $w_\BG(t)$ can be modeled as a sequence of 
discrete jumps:
\beq
  w_{\rm BG}(t)~=~ w_{\rm BG}^{(0)} + \sum_{\{i\}} 
    \Delta^{(i)} \,\Theta(t-t_i)~
\eeq
where $\{i\}$ labels an arbitrary collection of times $t_i$ at which $w_{\rm BG}$ 
suddenly changes by an amount $\Delta^{(i)}$.  In the infinitesimal limit of such 
jumps, one can obtain a continuous time dependence.

We show two examples of such jumps in Fig.~\ref{fig:w_track}.
In each of these examples, $w_{\rm BG}$ experiences two instantaneous 
jumps with $\abs{\Delta^{(i)}}=0.1$.  In the case represented by the solid 
red curve, $w_{\rm BG}$ increases twice.  By contrast, in the case represented 
by the dashed red curve, $w_{\rm BG}$ increases once and then drops back to 
its original value.

In Fig.~\ref{fig:w_track}, we also indicate  the response of a system with 
$\overline{w}= - 0.5$ to these sequences of jumps.  For each sequence of jumps, 
we see that the equation-of-state parameter $\expt{w}$ for our dynamical-scalar 
system (represented by the corresponding blue curve) always attempts to match these 
discrete changes in $w_{\rm BG}$, and even occasionally has enough time to 
achieve a short-lived stasis with $\langle w\rangle \to \overline{w}'=w_\BG$
until $w_{\rm BG}$ changes again. 
We also observe from Fig.~\ref{fig:w_track} that this tracking is not 
instantaneous.  In particular, although any stasis that has already been achieved 
is immediately destabilized when $w_{\rm BG}$ changes, it takes a non-zero amount 
of time for the system to realize the new stasis for which the new value of 
$\expt{w}$ matches the new value of $w_{\rm BG}$.  It is nevertheless intriguing 
to see that the tower is capable of adjusting its internal dynamics spontaneously 
in order to follow the change in the external background.

Such tracking behavior is not unexpected.  After all, our stasis solution is 
a global attractor.  As a result, any change in $w_{\rm BG}$ simply amounts 
to moving the location of the attractor within the phase space.  Of course, 
the trajectory lines that lead towards the original attractor are different from the 
trajectory lines that lead towards the new attractor.  However, any point on an original 
trajectory line is also a point on a new trajectory line.  Thus, at the moment 
when $w_\BG$ changes, our system simply begins to evolve along the new trajectory 
line rather the previous one.  For this reason, even if the background 
equation-of-state parameter varies continuously, the tower will always evolve 
in such a manner as to track $w_{\rm BG}$.  Of course, a perfect 
slow-roll/oscillatory-component stasis with $\langle w\rangle= w_\BG$ can 
only be expected once $w_{\rm BG}$ stabilizes to a constant.

There is also another potentially interesting aspect of the tracking behavior we have found.
Not only does our stasis solution relentlessly track the background when 
$ w_\BG< \overline{w}$, but it also {\it ceases}\/ to track the background if $w_\BG$ grows 
too large and exceeds $\overline{w}$.   Thus we have at our disposal not only a mechanism 
for tracking, but also a mechanism for {\it decoupling}\/ from tracking!  It is therefore 
possible to arrange things so that our stasis system not only tracks the background for a 
while but also later  decouples from it.  This feature can  accommodate a broader scope of 
expansion histories than possible in the traditional scenarios.

\FloatBarrier
\section{Towards a stasis-induced inflation}
\label{sec:inflation}

The results derived in Sects.~\ref{sec:ScalarTower} and \ref{sec:tracker} 
suggest that stasis could be the foundation of a deep connection between 
cosmology and particle physics.  Such a connection could then open the door 
to many new ways of thinking about the physics of the early universe.
Along these lines, one of the most exciting ideas that emerges from this 
work is the possibility that the stasis phenomenon we have discussed in this 
paper might serve as the foundation for a new approach to cosmological 
inflation.  In this section, we shall therefore outline some speculative 
thoughts concerning this possibility.

It is not hard to see that the stasis phenomenon could provide the underpinning 
of a possible new way of realizing cosmic inflation.  After all, as we have demonstrated, 
our system of rolling scalars can give rise to a stasis epoch during which the abundances 
of matter and vacuum energy remain fixed and in which the universe expands with 
an equation of state within the range $-1 < \overline{w} <0$.   Thus, if our 
initial conditions are such that $\overline{w}< -1/3$ (so that the universe 
experiences an accelerated expansion with $\ddot{a}>0$ during stasis), and if 
this stasis is maintained for a sufficient number of $e$-folds, we will have 
produced an epoch of accelerated expansion which could potentially explain 
the extraordinary flatness and homogeneity of our universe.  
This could then serve as the basis of a new model of cosmic inflation.

We shall refer to this intriguing possibility as a stasis-induced inflation, 
or simply ``Stasis Inflation"\/.  As it turns out, Stasis Inflation has a 
number of interesting features which merit further exploration.

First of all, Stasis Inflation does not require a complicated scalar potential.  Moreover, 
it does not rely on non-minimal coupling structures between the scalar sector 
and gravity, unlike many of the inflationary scenarios that have been proposed in order to
accommodate the most recent CMB measurements~\cite{BICEP:2021xfz, Planck:2018jri}. 
Indeed, the dynamics which gives rise to the accelerated expansion in Stasis Inflation does 
not follow primarily from the shape of the potential but rather from the structure of 
the underlying {\it particle physics}\/.  Moreover, in cases wherein the $\phi_\ell$ are the 
KK modes of a higher-dimensional scalar field, the mass spectrum of such states  
is primarily a reflection of the compactification geometry. 

Second, we observe that any equation-of-state parameter $\overline{w} <-1/3$ for the stasis 
sector can be realized within our Stasis Inflation framework.   Thus, it is relatively
straightforward to achieve an epoch of accelerated cosmological expansion within 
this scenario, and we are not restricted to having $\overline{w}\approx -1$.
Indeed, the ability of our stasis to ``track'' (and potentially later even decouple from) a time-dependent background may allow the value of $\overline{w}$ to vary throughout the inflationary epoch.  This in turn  provides us with a handle that might allow us to 
``dump'' unwanted relics (allowing them to inflate away) while simultaneously preserving other relics that might be more desirable.

As an example of this phenomenon, one could imagine inflating away cosmic strings 
(which have $w= -1/3$) while simultaneously preserving domain or bubble walls (which have $w=-2/3$) 
so long as  $\overline{w}'$ lies between these two values.  One could then imagine that 
$\overline{w}'$ drops further at a later time, inducing the domain walls to inflate away.   
In this way we would have not only arranged for both  cosmic strings and domain walls 
to inflate away but also separated the {\it times}\/ at which these processes occur within the 
inflationary epoch.  This in turn modifies the manner in which the domain walls associated with 
bubbles of different sizes exit the horizon, producing results which differ from what is possible 
in traditional inflation scenarios.  Likewise, this also modifies the manner in which a population 
of bubbles evolves once they subsequently re-enter the horizon.  Moreover, in cosmologies of this 
sort, the changing equation of state of the universe during the inflationary epoch modifies the 
manner in which density perturbations with different wavenumbers exit the horizon.  This can lead 
to the presence of non-trivial features in the spectrum of density perturbations --- features which 
could leave characteristic imprints in the CMB.~

Third, within the most natural realizations of Stasis Inflation the number of 
$e$-folds of inflation is no longer a free parameter but is directly related 
to the hierarchies between particle-physics scales.  For example, for scenarios 
in which the $\phi_\ell$ tower consists of the KK excitations of a higher-dimensional 
scalar field, the number of such states will generically scale as 
$N\sim (R M_{\rm UV})^n$  where $R$ schematically denotes a compactification radius, 
where $M_{\rm UV}$ denotes a UV cutoff such as the string scale or the  Planck scale, 
and where $n$ is the number of compactified dimensions.  We thus see that the number 
of states in the tower --- and thus the duration of the stasis epoch or equivalently 
the number of $e$-folds of cosmic inflation 
produced --- is directly connected to the hierarchy between $R^{-1}$ and $M_{\rm UV}$.  
Taking $R^{-1}\sim {\cal O}({\rm TeV})$ and $M_{\rm UV}\sim {\cal O}(M_{\rm Planck})$ ---
as is typical in theories involving large extra dimensions --- we see that this 
hierarchy can be significant.  Such models would then lead us to conclude that the 
universe is large simply because the Planck/TeV hierarchy is big!  This thereby 
provides a novel connection between two large numbers in physics.

Fourth, the Stasis Inflation scenario has a natural graceful exit.   
Indeed, the stasis epoch ends when the transitions from (overdamped) 
vacuum energy to (underdamped) matter have reached the bottom of the 
$\phi_\ell$ tower.  As we have seen, this sort of exit is a general feature 
of {\it all}\/ stasis epochs, and as such a graceful exit from accelerated 
expansion is already an inherent part of the Stasis Inflation scenario.

But finally --- and perhaps most importantly --- Stasis Inflation behaves differently 
than ordinary inflation in terms of its effects on the abundances of vacuum energy, 
matter, radiation, and potentially even other energy components in the universe.   
Normally, during traditional inflationary epochs, the universe rapidly becomes dominated 
by vacuum energy, and all other energy components that might have existed at the start 
of inflation will inflate away, with abundances that fall to zero.  This is ultimately 
why an epoch of reheating is generally required after traditional inflation.   
However, with Stasis Inflation, the situation is different: a non-zero matter abundance 
can be carried along throughout the inflationary epoch without exhibiting any 
reduction, even though the universe is undergoing an accelerated expansion with 
$\overline{w}< -1/3$.   This is ultimately because vacuum energy and matter 
together play a crucial role in sustaining
the inflationary stasis that drives the cosmological inflation.  
Moreover, if we further allow our scalar fields to decay to radiation, as discussed 
in Ref.~\cite{Dienes:2023ziv}, we can even achieve an inflationary {\it triple}\/ stasis 
involving not only vacuum energy and matter but also radiation.  Thus the abundances 
of vacuum energy, matter, and radiation can all be sustained across the inflationary epoch.  
This has the potential to significantly change the conditions needed for any subsequent 
reheating.
      
In this section we have provided only a rough qualitative sketch of a possible Stasis 
Inflation scenario.  Much more work is needed in order to determine whether such a 
scenario is phenomenologically viable.  For example,  Stasis Inflation must be shown  
to generate the correct power spectrum for scalar perturbations while satisfying 
current bounds on tensor perturbations~\cite{BICEP:2021xfz, Planck:2018jri}.  Similarly, 
one must examine the generation of non-Gaussianities and isocurvature perturbations 
within such scenarios and determine whether the results are consistent with current 
observational   constraints~\cite{Planck:2019kim,Planck:2018jri}.  Stasis Inflation 
nevertheless remains an interesting possibility worthy of further exploration.


\FloatBarrier
\section{Discussion and conclusions\label{sec:Conclusions}}


Towers comprising large numbers of scalar fields are a common feature of many BSM 
scenarios, including theories with extra spacetime dimensions and string theory.
Moreover, the homogeneous zero-mode field value associated with each of these fields 
transitions dynamically from an overdamped phase exhibitng slow-roll behavior to an 
underdamped phase exhibiting rapid oscillations.  In this paper, we have examined 
the conditions under which the full dynamics of such fields across the tower 
can give rise to an epoch of cosmic stasis between the collective slow-roll and 
oscillatory abundances associated with these two phases. 

In the simplest case one might consider --- that in which no additional 
cosmological energy components are present and in which 
the masses and initial abundances of these scalars across the 
tower follow the scaling relation in Eq.~(\ref{eq:initial_abundance})  --- 
we found  that a cosmological stasis can develop during which these two collective 
abundances $\Omega_{\rm SR}$ and $\Omega_{\rm osc}$ each remain constant despite 
cosmic expansion across many $e$-folds.  Indeed, a stasis of this sort arises 
generically in any such system, provided that our mass and abundance 
scaling exponents $\delta$ and $\alpha$ satisfy $\alpha+1/\delta=2$ and provided that the 
density of states per unit mass within the tower is sufficiently large that 
the mass spectrum may reliably be approximated as a continuum.   
Moreover, we also demonstrated that in such circumstances,
the stasis state is actually a cosmological attractor.
Depending on initial conditions, the ultimate stasis abundances 
$\barOmega_{\rm SR}$ and $\barOmega_{\rm osc}$ can take any value within the 
ranges $0 < \barOmega_{\rm SR} < 1$ and 
$0 < \barOmega_{\rm osc} < 1$, with $\barOmega_\SR + \barOmega_\osc =1$.  
As a result, the effective equation-of-state parameter for the universe 
as a whole during stasis can take any value within the range $-1 < \barw < 0$. 

We also considered how this picture changes in the presence of an additional 
background energy component beyond the scalar tower --- a component that we take 
to be a perfect fluid with a constant equation-of-state parameter $w_{\rm BG}$ and 
arbitrary initial abundance $\Omega_{\rm BG}^{(0)}$.  We found that a stasis always 
emerges in this case as well.  However, we found that the properties of this stasis 
depend on the relative sizes between $w_\BG$ and $\overline{w}$, where $\overline{w}$ is 
the effective equation-of-state parameter of the stasis that would have emerged in
the absence of this additional cosmological energy component. 
When $w_{\rm BG} > \barw$, we found that the background energy 
density $\rho_{\rm BG}$ decreases more rapidly with cosmic expansion than 
does the energy density of the tower.  Thus the properties of the resulting 
asymptotic stasis are unaffected by the presence of the background, and 
our tower gives rise to the same stasis as before.   By contrast, in cases 
in which $w_{\rm BG} < \barw$, we found that the tower gives rise to an 
entirely new stasis --- a {\it tracking}\/ stasis  --- in which the new 
equation-of-state parameter $\barw'$ evolves toward the value $w_{\rm BG}$.
Indeed, this is true regardless of the initial background abundance 
$\Omega_{\rm BG}^{(0)}$. 

Finally, we speculated that these ideas might form the basis of a new approach 
towards understanding cosmic inflation.   Indeed, a stasis epoch of this sort 
with $\overline{w} < -1/3$ exhibits accelerated expansion, and could potentially 
solve the flatness and hierarchy problems.   Moreover, as we discussed, 
this sort of ``Stasis Inflation'' has a number of intriguing and potentially 
beneficial properties not shared by traditional models of inflation.   Of course, 
a more detailed examination of this idea is necessary before any conclusions 
concerning its viability can be drawn.    

A few comments are in order.  First, since the particular form of stasis that we have 
examined in this paper arises from the collective dynamics of a large number of 
scalar fields whose masses and abundances exhibit particular scaling behaviors, 
it is important to consider how these scaling behaviors might arise in 
actual models of particle physics.  Fortunately, the mass spectrum that we 
considered in this paper --- a spectrum characterized by a scaling exponent 
$\delta$, a mass-splitting parameter $\Delta m$, and a ground-state mass shift 
$m_0$ --- is a fairly generic one in many extensions of the Standard Model (SM), 
including those extensions involving extra compact spacetime dimensions.
Moreover, the spectrum of initial abundances that  we  considered is one in 
which the $\Omega_{\ell}^{(0)}$ scale with $m_\ell$ according to a power law.  
As we have discussed, this too is a fairly generic result emerging from many 
different types of production mechanisms.   

Given the requirements of stasis, we found that the spectrum of initial field 
displacements $\phi_\ell^{(0)}$ in our model must scale as 
$\phi_\ell^{(0)}\sim \ell^{-1/2}$, assuming all such fields start from rest.
While such a spectrum of initial displacements is in principle achievable in 
scenarios in which the $\phi_\ell$ are the KK excitations of a higher-dimensional 
scalar field, it would be interesting 
to explore how such a spectrum might emerge within the framework of a more 
fully developed model of higher-dimensional physics.  Of course, explicit 
model constructions exist in the 
literature~\cite{Dienes:2011ja,Dienes:2011sa,Dienes:2012jb} wherein the 
initial abundances $\Omega_\ell^{(0)}$ and masses $m_\ell$ obey the same 
scaling relations across a tower of axion-like particles  as we have assumed 
here, with scaling exponents $\alpha$ and $\delta$ respectively. Those models 
were developed in order to address the dark-matter problem, and there is even 
a partial overlap in the $(\alpha,\delta)$ parameter space between what is 
required for those models and what we require for stasis.   However, it 
still remains to construct explicit particle-physics models within that 
overlap region. 

For simplicity, we have also assumed throughout this paper that while the initial 
displacements for our $\phi_\ell$ fields can be sizable, the corresponding initial 
velocities $\dot{\phi}_\ell^{(0)}$ are sufficiently small that their impact on the 
subsequent evolution of the system can be neglected.  There are a number of ways in 
which such initial conditions for the $\dot{\phi}_\ell^{(0)}$ can be achieved.  
One possibility, which is discussed in Ref.~\cite{Dienes:1999gw}, arises within
the context of theories with extra spacetime dimensions in which the SM is localized 
on a four-dimensional brane, while an additional scalar field neutral under the SM 
gauge group propagates freely within the higher-dimensional bulk.  The KK modes of this 
bulk scalar play the role of the $\phi_\ell$ in this context.  At early times, 
the mass eigenstates of the theory are simply the KK-number eigenstates.  
One natural set of initial conditions for the zero-modes of these scalars, which 
follows from positing that by some early time they have all effectively settled into 
their potential minima, is $\phi_\ell\approx0$ and $\dot{\phi}_\ell \approx 0$ for 
all of the $\phi_\ell$ except $\phi_0$, which is massless and can therefore acquire a 
misaligned background value.  If additional terms in the mass matrix for the 
$\phi_\ell$ are generated dynamically at some later time --- for example, as the 
consequence of a second-order phase transition --- the mass eigenstates of the theory
are subsequently admixtures of the KK-number eigenstates.  These late-time eigenstates 
acquire non-zero field displacements (but negligible field velocities) via their mixing 
with the KK zero-mode.

Another possibility is that the $\dot{\phi}_\ell$ are initially non-negligible at
very early times.  However, the kinetic-energy density $\rho_\ell^{({\rm KE})}$ associated with 
each of these field velocities damps away rapidly in comparison with the corresponding potential 
energy.  As a result, the universe will ultimately tend toward a state in which all of the 
$\phi_\ell$ are either slowly rolling or oscillating around their potential minima.  This suppression 
of kinetic-energy density occurs generically, regardless of the effective equation-of-state 
for the universe prior to the onset of stasis, though the rate at which the 
$\rho_\ell^{({\rm KE})}$ are suppressed of course depends on this equation-of-state parameter.
One possibility which would result in a particularly efficient suppression of the 
$\rho_\ell^{({\rm KE})}$ involves positing that a brief auxiliary period of accelerated cosmic 
expansion took place before stasis began.  Such a period of accelerated expansion could potentially 
even be driven by the $\phi_\ell$ themselves, provided that $H^{(0)}/m_{N-1}$ is sufficiently large.  
During this period, the $\rho_\ell^{({\rm KE})}$ 
would decrease exponentially with $t$.  We emphasize this period need not last more than a few 
$e$-folds in order to establish an appropriate set of initial conditions for the $\phi_\ell$, 
and therefore would not usurp the role of the subsequent stasis epoch in providing a solution 
to the horizon problem.

Second, spatially homogeneous scalar-field configurations are generically
unstable due to parametric resonances associated with self-interaction terms in the 
scalar potential (if such terms are present)~\cite{Kofman:1994rk,Shtanov:1994ce,Kofman:1997yn} 
or to gravitational instabilities~\cite{Jedamzik:2010dq,Easther:2010mr,Musoke:2019ima}.  As
a result, the homogeneous zero-mode configurations of our $\phi_\ell$ are expected to fragment 
over time into a collection of spatially localized, gravitationally bound structures or ``lumps.''  
However, the fragmentation of the initially homogeneous field configurations of our $\phi_\ell$ 
should not have a significant impact on the dynamics which gives rise to stasis.  Indeed, this 
fragmentation, which for the purely quadratic form of $V_\ell$ we are considering in this 
paper is due to gravitational instabilities, occurs only after the field in question begins
oscillating~\cite{Jedamzik:2010dq}.  Moreover, for this form of $V_\ell$, the lumps
into which each $\phi_\ell$ condensate fragments behave like massive matter.
Thus, both before and after fragmentation, each $\phi_\ell$ which satisfies $3H(t) < 2m_\ell$ 
at a given time $t$ and therefore contributes to $\Omega_{\rm osc}$ has an equation-state 
parameter $w_\ell \approx 0$.  Fragmentation therefore has no fundamental impact on the 
cosmological dynamics which gives rise to stasis in this case, irrespective of the value of 
$\barw$.  This conclusion still holds even if the $V_\ell$ potentials are not purely quadratic, but 
include additional terms which are higher-order in $\phi_\ell$, provided that the anharmonicities are 
sufficiently small.  

For other forms of $V_\ell$ (\eg, for quartic potentials), the gravitationally bound structures 
into which the $\phi_\ell$ condensates fragment do not necessarily behave like massive matter.  
That said, in the inflationary regime in which $\barw \leq -1/3$, the energy density associated 
with these structures rapidly redshifts away during stasis.  Since each individual field only
contributes significantly to $\Omega_{\rm osc}$ for a brief period after it has begun oscillating,
and since fragmentation of each $\phi_\ell$ condensate in this case likewise occurs only once the 
field has begun oscillating, the impact of fragmentation on a stasis with such a value of $\barw$ 
may also be negligible in many cases, depending on the rate at which $\phi_\ell$ particles are 
produced via parametric resonance.  We leave a detailed study of this dynamics for future work.

Third, while we have focused in this paper on the case in which the potentials
for the $\phi_\ell$ are quadratic, there are also other forms for the potential which are
of interest from a model-building perspective.  One of these is the form
\begin{equation}
    V(\phi_\ell) ~=~ \sum_i \,\Lambda_i \,\exp 
      \left(\sum_\ell \alpha_{i\ell}\, \phi_\ell\right)~,~~
  \label{eq:VScalingSol}
\end{equation}
where the $\Lambda_i$ and $\alpha_{i\ell}$ are model-dependent constants.
Potentials of this form arise in the low-energy limits of various string compactifications 
and are of significant interest because they can give rise to so-called scaling 
cosmologies~\cite{Shiu:2023nph,Shiu:2023fhb} (see also Ref.~\cite{Calderon-Infante:2022nxb}) 
in which the scale factor $a(t)$ evolves with time $t$ according to a power law.  While such 
potentials lack stable minima, it is nevertheless conceivable that they could also give rise to 
a stasis epoch.  Indeed, even in scenarios involving a single scalar field with an exponential 
potential~\cite{Copeland:1997et}, one finds that the equation-of-state parameter 
$w$ for the single scalar field $\phi$ is approximately $w(t) \approx -1$ at 
early times. However, since there is no potential minimum, this field gathers 
kinetic energy as it rolls, and as a result $w(t)$ increases.
It would be interesting to investigate how this behavior is modified in a multi-field
context and to examine the extent to which the dynamical 
evolution from smaller to larger values of $w_\ell(t)$ can compensate for the effect of 
cosmological expansion, thereby potentially giving rise to a stasis epoch.  
We leave this possibility for future work.

Fourth and finally, we emphasize again that the ideas in this paper could serve as the 
foundation of a deep connection between cosmology and particle physics and thereby open 
the door to many exciting phenomenological implications of cosmic stasis.
For example, as discussed in Sect.~\ref{sec:inflation}, a stasis epoch with 
$\barw < -1/3$ persisting for $\mathcal{N}_s \gtrsim 60$ $e$-folds of cosmic expansion can in 
principle constitute a solution to the flatness and horizon problems.  If indeed
viable models of Stasis Inflation could be developed along these lines, such models would 
almost certainly give rise to distinctive power spectra of primordial scalar and tensor 
perturbations.  This opens the possibility that evidence for Stasis Inflation could 
potentially be extracted from observations of the CMB and/or the stochastic 
gravitational-wave background.  
Another possibility is that a stasis involving dynamical scalars could occur
much later in the cosmological timeline.  In particular, it would be interesting to consider 
the possibility that the slowly rolling $\phi_\ell$ could collectively constitute 
the dark energy which drives the accelerated expansion that we observe at the present time, 
while the oscillatory $\phi_\ell$ could collectively constitute the dark matter. 
Of course, the presence of large numbers of extremely light axion-like scalars
presents a number of model-building challenges, including those imposed by 
constraints on supernova energy loss~\cite{Raffelt:1987yt,Olive:2007aj}, 
E\"{o}tv\"{o}s-type experiments~\cite{Chen:2014oda,Tan:2020vpf}, searches for 
frequency variation in atomic clocks~\cite{Arvanitaki:2014faa}, black-hole 
superradiance considerations~\cite{Arvanitaki:2010sy,Brito:2015oca,Cardoso:2018tly,
Stott:2018opm,Stott:2020gjj,Mehta:2020kwu}, data from pulsar-timing 
arrays~\cite{Blas:2016ddr,Kaplan:2022lmz}, and the inverse correlation which exists 
between the dark-energy equation of state and the present-day value of $H$ in combined 
fits to CMB and supernova data~\cite{Zhao:2017cud,Vagnozzi:2018jhn,Vagnozzi:2019ezj,
Banerjee:2020xcn,Lee:2022cyh}. That said, if one were to construct
a phenomenologically viable model of dark energy along these lines, it would go a long
way toward addressing the cosmic coincidence problem.  
This topic is therefore worthy of further exploration.


\begin{acknowledgments}

The research activities of KRD are supported 
in part by the U.S.\ Department of Energy under Grant DE-FG02-13ER41976 / DE-SC0009913, and 
also by the U.S.\ National Science Foundation through its employee IR/D program.
 The work of LH is supported by the STFC (grant No. ST/X000753/1).  
The work of FH is supported in part by the Israel Science Foundation grant
1784/20, and by MINERVA grant 714123.
The work of TMPT is supported in part by the U.S.\ National Science Foundation under 
Grant PHY-2210283.  The research activities of BT are supported in part by the 
U.S.\ National Science Foundation under Grants PHY-2014104 and PHY-2310622. 
BT also wishes to acknowledge the hospitality of the Kavli Institute for Theoretical 
Physics (KITP), which is supported in part by the U.S.\ National Science Foundation under 
Grant PHY-2309135.  The opinions and conclusions
expressed herein are those of the authors, and do not represent any funding agencies. 

\end{acknowledgments}

\vfill\eject

\bigskip
\bigskip

\bibliography{TheLiterature2}

\end{document}